\documentclass[usenatbib]{mnras}

\usepackage{float}
\usepackage{xcolor}

\usepackage{graphicx}	
\usepackage{amsmath}	
\usepackage{amssymb}	
\usepackage{tabularx}
\usepackage{soul}
\usepackage{bm}
\usepackage{xspace}
\usepackage{subcaption}

\title[IPTA DR2 GWB]{The International Pulsar Timing Array second data release: Search for an isotropic Gravitational Wave Background }

\author[IPTA]{\parbox{\textwidth}{\large
J.~Antoniadis,$^{1,2,3}$
Z.~Arzoumanian,$^{4}$
S.~Babak,$^{5,6}$
M.~Bailes,$^{7,8}$
A.-S.~Bak~Nielsen,$^{2,9}$
P.~T.~Baker,$^{10}$
C.~G.~Bassa,$^{11}$
B.~B\'{e}csy,$^{12}$
A.~Berthereau,$^{13,14}$
M.~Bonetti,$^{15,16}$
% ...
}
\vspace{0.4cm} \\ 
Affiliations are at the end of the paper
}

\date{Accepted XXX. Received YYY; in original form ZZZ}

\pubyear{2021}

\begin{document}
\label{firstpage}
\pagerange{\pageref{firstpage}--\pageref{lastpage}}
\maketitle

\begin{abstract}

We searched for an isotropic stochastic gravitational wave background in the second data release of the International Pulsar Timing Array, a global collaboration synthesizing decadal-length pulsar-timing campaigns in North America, Europe, and Australia.
In our reference search for a power law strain spectrum of the form $h_c = A(f/1\,\mathrm{yr}^{-1})^{\alpha}$, we found strong evidence for a spectrally-similar low-frequency stochastic process of amplitude $A = 3.8^{+6.3}_{-2.5}\times10^{-15}$ and spectral index $\alpha = -0.5 \pm 0.5$, where the uncertainties represent 95\% credible regions,
using information from the auto- and cross-correlation terms between the pulsars in the array. For a spectral index of $\alpha = -2/3$, as expected from a population of inspiralling supermassive black hole binaries, the recovered amplitude is $A = 2.8^{+1.2}_{-0.8}\times10^{-15}$.
Nonetheless, no significant evidence of the Hellings-Downs correlations that would indicate a gravitational-wave origin was found.
We also analyzed the constituent data from the individual pulsar timing arrays in a consistent way, and clearly demonstrate that the combined international data set is more sensitive. Furthermore, we demonstrate that this combined data set produces comparable constraints to recent single-array data sets which have more data than the constituent parts of the combination.
Future international data releases will deliver increased sensitivity to gravitational wave radiation, and significantly increase the detection probability.

\end{abstract}


\section{Introduction} \label{sec:intro}

Inspiralling supermassive black hole binaries (SMBHBs) with masses larger than $10^7 M_\odot$ are expected to generate the strongest gravitational-wave (GW) signals in the Universe. The incoherent superposition of all of these inspiralling SMBHBs should generate a stochastic GW background (GWB) that is the strongest in the nanohertz frequency band \citep[e.g.,][]{rr95a,jb03,svc08,bs+2019}.
Other sources that could also produce a stochastic background in the nanohertz band are cosmic strings \citep[e.g.,][]{GWB_strings_1}, cosmological phase transitions, and a primordial background produced by quantum fluctuations in the gravitational field in the early universe \citep[e.g.,][]{PGW_1,PGW_2}.
For comparison, the Laser Interferometer Gravitational-Wave Observatory (LIGO) and the Virgo Collaboration, which are terrestrial GW detectors and have detected GWs from merging stellar mass black holes and neutron stars \citep[e.g.,][]{gwtc-1, gwtc-2}, are only sensitive to GW signals that are ten orders of magnitude higher in frequency than PTAs.

A nanohertz GWB can be detected using a precisely timed ensemble of millisecond pulsars~\citep{1978sazhin,det79}, called a pulsar timing array~\citep[PTA,][]{1990FosterBacker}. 
The GWs distort the space-time between the Earth and pulsars, changing their proper distance, thereby leading to a measurable deviation of the pulsar pulse arrival times. Since such effects cannot be detected with confidence using only one pulsar, PTAs leverage the imprint of \textit{spatially-correlated} timing deviations between pulsars which are separated by kiloparsec distances across the galaxy, yet are subject to the common influence of the GWB.  

An isotropic GWB manifests itself as a long timescale, low-frequency (or red) common signal across the pulsars in a PTA. This common signal is characterized by the common spectrum and the inter-pulsar spatial correlations. For an isotropic GWB these spatial correlations are unique, referred to as the \cite{HellingsDowns} (HD) correlations, and thus are considered to be the ``smoking gun'' signature for the presence of a GWB \citep[][]{thk+2018} in any PTA data set.
The spectral amplitude of this common signal is determined by the characteristic strain, $h_c(f)$, of the GWB, which itself is a function of the physics sourcing the GWB (e.g., SMBHB masses, merger timescale, and number density) \citep[e.g.,][]{2013CQGra..30v4014S,kbh+2017,csc2019}. Thus, precise spectral characterization of the GWB will allow us to extract the underlying astrophysics of the background, as well as distinguish between different sources of the GWB \citep[e.g.,][]{Pol:2020igl}.

The ability to detect GWs relies on, among other things, the number of pulsars available to cross-correlate in GWB searches, and on the length of each pulsar data set \citep[][]{sej+13}. Improvements in both of these parameters increases the detection significance of the GWB signal which in turn allows for better constraints on the parameters of the GWB spectrum \citep[][]{Pol:2020igl}.
Hence, international efforts spanning decades from the European Pulsar Timing Array \cite[EPTA,][]{2016MNRAS.458.3341D}, North American Nanohertz Observatory for Gravitational Waves  \cite[NANOGrav,][]{2016ApJ...821...13A}, and the Parkes Pulsar Timing array \cite[PPTA,][]{mhb+13}, as well as newer PTAs such as the Indian Pulsar Timing Array \citep[InPTA,][]{InPTA}, Chinese Pulsar Timing Array \citep[CPTA,][]{CPTA}, and with the MeerKAT Interferometer in South Africa \citep[][]{SAPTA} share and combine their data to form the International PTA \citep[IPTA,][]{ipta2010}.

In this spirit of international collaboration the IPTA has produced two data sets to date. The first IPTA data release \citep[DR1,][]{vlh+16} consisted of 44 millisecond pulsars and yielded no conclusive detection of a GWB. The second IPTA data release \citep[DR2,][]{pdd+19} consists of 65 pulsars and is the focus of this analysis. The pulsars in DR2 have data sets spanning $0.5-30$ years. For the first time, we process the data subsets from each individual PTA and search for a GWB in a self-consistent way, thus enabling us to make a fair comparison of respective PTA constraints. 

Recently, a spatially uncorrelated (pulsar-weighted-average) spectrally similar common process or common-spectrum process (CP) was detected in the NANOGrav 12.5-year data set \citep{abb+2020}, the second data release of the Parkes Pulsar Timing Array~\citep{ppta_dr2_gwb}, and the EPTA six-pulsar data set of the second data release~\citep{epta_dr2_gwb}.
The process is modeled as an additional time-correlated term with the same power spectrum in all of the pulsars.
However, there is little evidence to support the existence of spatial HD correlations in any of these data sets. We compare the IPTA DR2 constraints on the GWB with those obtained from these analyses.

The paper is organized as follows: in Section \ref{sec:DR2} we give an overview of the second IPTA data release, hereafter referred to as DR2. We describe our data analysis methods in Section \ref{sec:DAmethods}, and give our results in Section \ref{sec:results}. Caveats and implications of our analysis and results are discussed in Section \ref{sec:discuss}, including the astrophysical interpretation of a potential GWB. The conclusion is given in Section \ref{sec:conclusion}.

\section{IPTA Data Release 2}
\label{sec:DR2}

IPTA DR2 includes a combination of timing data from the following individual PTA data releases: the EPTA data release 1.0 \citep{2016MNRAS.458.3341D}, the NANOGrav 9-year data set \citep{abb+15}, and the PPTA first data release \citep{mhb+13} and its extended version \citep{rhc+16}. The EPTA data set includes high-precision timing data from 42 MSPs obtained with the largest radio telescopes in Europe -- Effelsberg telescope, Lovell telescope, Nan\c{c}ay telescope, and Westerbork Synthesis telescope – covering data from 1996 to 2015 with a time baseline between 7--18 yr. In addition to these data, archival timing data of PSR J1939$+$2134 since 1994 was included. The NANOGrav 9-year data set includes high-precision timing observations from 37 MSPs obtained with the Robert C. Byrd Green Bank Telescope and the Arecibo telescope, spanning a time baseline between 0.6--9.2 yr, covering the data from 2004 to 2013. In addition, the long-term timing data of PSR J1713$+$0747 from \citet{zsd+15} and the data of PSRs J1857$+$0943 and J1939$+$2134 from 1984 through 1992 \citep{ktr94} were included. The PPTA data set includes high-precision timing observations from 20 MSPs obtained with the Parkes radio telescope (also known as {\em Murriyang}) from 2004 to 2011. IPTA DR2 also included single frequency band (1.4\,GHz/L-band) Parkes telescope legacy data obtained since 1994. The additional 3.0\,GHz timing data reported in \citet{srl+15} for PSRs J0437$-$4715, J1744$-$1134, J1713$+$0747, and J1909$-$3744 were also included in the data set. In total, the timing data from 65 MSPs were included in IPTA DR2, which has 21 more source than the IPTA DR1 \citep{vlh+16}. There are 27 and 7 MSPs in IPTA DR2 with a timing baseline $>$10 yr and $>$20 yr, respectively. All pulsars were observed at multiple frequencies. All EPTA and PPTA observations were averaged in time and frequency to obtain a single time-of-arrival (TOA) for each receiver and observation. The NANOGrav observations were averaged in time and included sub-band information, i.e., averaged in frequency to maintain a frequency resolution ranging from 1.5 to 12.5~MHz depending on the receiver and backend instrument combination, resulted in a single TOA for each frequency channel. More details about the constituent PTA data sets can be found in \citet{pdd+19}.

The different data sets for a given pulsar in IPTA DR2 were combined by fitting for time offsets, referred to as JUMPs, in the timing model to account for any systematic delays between data sets. The highest weighted data set with the lowest sum of TOA uncertainties was used as the reference data set in this process. The timing models of pulsars included astrometric parameters, rotational frequency information, dispersion measure information, and Keplerian and Post-Keplerian parameters if the pulsar is in a binary system. For NANOGrav observed pulsars, ``FD'' parameters were included to minimize the effect of frequency-dependent profile variations of pulsars \citep[see][]{abb+15}. IPTA DR2 produced two data set versions depending on different methods of handling the dispersion measures (DM) variations of pulsars over time \citep[VersionA and VersionB -- see][for details]{pdd+19}. In VersionA, the DM variations of pulsars were determined using DMMODEL described in \citet{kcs+13} and the noise parameters for different data sets were directly taken from their original data releases. In VersionB, the DM variations were modeled using the first two time derivatives of the DM and a time-correlated stochastic DM process in the timing model. The noise parameters were also re-estimated based on the new IPTA data combination in this version. We use VersionB for this work.

\section{Data-analysis Methods}
\label{sec:DAmethods}

In this work we follow the conventions established by other pulsar timing array data analyses~\citep[i.e.,][]{2016ApJ...821...13A,abb+2020,ltm+15,ppta_dr2_gwb}.
The multivariate Gaussian likelihood $\mathcal{L}(\bm{\delta t} | \bm{\theta})$ is employed to model noise and signal contributions, parametrised by $\bm{\theta}$, to the observed timing residuals.
Our likelihood was of the same form as other PTA analyses, \cite[e.g.,][]{abb+15}.
We used \texttt{enterprise} \citep{enterprise} to evaluate the likelihood and priors, and \texttt{PTMCMCSampler} \citep{ptmcmc} to perform a Markov chain Monte Carlo (MCMC) simulation, drawing samples from the posterior probability distribution.
Model selection was performed via the product-space sampling method~\citep{carlinchib1995,hhh+2016}.
Additionally, we used the Savage-Dickey approximation to the Bayesian evidence ratio when appropriate.

\subsection{Noise models}
\label{sec:DAmethods.noise}

For each pulsar, we modeled the TOAs with a combination of four processes: the timing model, white noise, intrinsic red noise, and DM variations.
Deterministic contributions from the timing model, described in Section~\ref{sec:DR2}, were analytically marginalised~\citep{vanhaasteren2009}.
The time-uncorrelated white noise was modeled with EFAC and EQUAD, and ECORR parameters for NANOGrav pulsars with their sub-banded TOAs \citep[definitions of EFAC, EQUAD, and ECORR can be found in, e.g.,][]{vlh+16,pdd+19}.
Every observing receiver and backend system combination is given its own set of white noise parameters.
The time-correlated red noise process \cite[e.g., pulsar spin noise,][]{sc10} and stochastic DM variations \cite[][]{kcs+13} were modeled as Fourier basis Gaussian processes.
In each case Fourier spectrum coefficients were modeled as power laws,
\begin{align}
    P_{\rm RN}(f) &= \frac{{A_{\rm RN}}^2}{12 \pi^2} f_\mathrm{yr}^{-3}  \left(\frac{f}{f_\mathrm{yr}}\right)^{-\gamma_{\rm RN}} \label{eqn:psd} \,, 
    \\
    P_{\rm DM}(f,\nu) &=\frac{{A_{\rm DM}}^2}{12 \pi^2} f_\mathrm{yr}^{-3}  \left(\frac{f}{f_\mathrm{yr}}\right)^{-\gamma_{\rm DM}} \left(\frac{1400 {\rm MHz}}{\nu}\right)^2 \,,
\end{align}
where $A$ is the power law amplitude, $\gamma$ its spectral index, and $f_\mathrm{yr} = 1\;\mathrm{yr}^{-1} \approx 3.17 \times 10^{-8}\;\mathrm{Hz}$.
The difference between these two processes lies in the radio frequency $\nu$ dependence.
Intrinsic red noise is achromatic, i.e., frequency independent, while DM variations follow a $\nu^{-2}$ dependence \citep[e.g.,][]{lsc+16}. 

Despite all MSPs in IPTA DR2 exhibiting high rotational stability, such that the marginalized timing model, red, and DM noise terms are in general sufficient, certain pulsars have been found to experience timing events that need to be included in their data model.
Of interest to this analysis is PSR J1713$+$0747, which was observed to experience multiple sudden drops in apparent DM with an exponential recovery~\citep{dfg+13,lam+2018,Goncharov:2020krd}.
Only the first such event lies within the timespan of the IPTA DR2 and was included as an additional deterministic term to the full noise model of PSR J1713$+$0747. The amplitude, epoch and recovery time scale of the DM exponential dip are sampled simultaneously with the pulsar red and DM noise terms.

\cite{lsc+16} found additional sources of red noise in IPTA DR1.
These include radio frequency band-dependent and observing system-dependent terms, which may affect measurements of $P_{\rm RN}$ and $P_{\rm DM}$, if not modeled.
It is possible that mismodeling these effects can bias recovery of the CP.
More prescriptive models for the CP should be less affected by this bias.
Recent PTA analyses have included more complex red noise and DM variation models, where different pulsars in the array use different models \citep[e.g.,][]{ng11cw,Goncharov:2020krd}.
In the name of computational efficiency we opted to use the same power law models for all pulsars except when absolutely necessary, as is the case for PSR J1713$+$0747.

IPTA DR2, being the combination of data from multiple telescopes and many observing systems, has larger model parameter space than its constituent data sets.
The large number of model parameters and TOAs increases the computational complexity of the analysis.
As we searched for long-term processes, such as the GWB, we limited our analysis to pulsars whose observation time exceeded 3 years. This reduced the number of pulsars from the full 65 in DR2 to 53.
Additionally, we fixed the white noise parameters (EFAC, EQUAD, and ECORR) to median aposteriori values from single pulsar analyses.
Both of these choices reduced the analysis parameter space to a more manageable size. 

\subsection{Common-spectrum process models}
\label{sec:DAmethods.commonprocess}
In addition to modeling noise intrinsic to the individual pulsars, 
we also include a red CP that is present in all of the pulsars.
The source of this process could be the GWB, or any other common noise that manifests itself in all pulsars, such as clock errors \citep{2016noiseEPTA,hgc+2020} or errors in the Roemer delays from Solar-system ephemeris (SSE) systematics \citep{thk+2018,vts+2020}.
The choice of red noise priors also affects the recovery of a CP due to covariance between pulsar intrinsic red noise and the CP \citep{Hazboun:2020b,ppta_dr2_gwb}.
Each of these effects can be distinguished by a unique pattern of spatial cross-correlations between pulsars.
The cross-power spectral density is defined as,
\begin{equation}
    S_{ab}(f) = \Gamma_{ab} P_{\rm CP}(f) \,,
    \label{eqn:crossspect}
\end{equation}
where $P_{\rm CP}$ is the common-spectrum process and $\Gamma_{ab}$ is the overlap reduction function (ORF) describing the inter-pulsar correlations.

For some analyses we did not account for any inter-pulsar correlations, taking $\Gamma_{ab}=\delta_{ab}$ to be the identity matrix.
For others, we also included different choices of non-diagonal ORFs, such as the quadrupolar \citet{HellingsDowns} correlations that describe a GWB, dipolar correlations associated with SSE errors, or monopolar correlations, $\Gamma_{ab}=1$, associated with clock errors.
In some cases we split the diagonal auto-correlation part of the ORF from the off-diagonal cross-correlation part, treating them as independent processes as a consistency check.
When modeling the CP using only the auto-correlations, it is possible to analyze the data from each pulsar independently, then recombine the results to achieve a joint posterior on the CP.
We refer to this as the \textit{factorized likelihood} approach.

We modeled the CP using a Fourier basis Gaussian process, using basis frequencies $f = 1/T, 2/T, \ldots$, 
where $T$ is the timespan between the earliest and latest observation in the data set.
We model the power spectrum of the CP as a power law using Eqn. \eqref{eqn:psd}, replacing the pulsar noise amplitude and spectral index with those from the common process $A_\mathrm{CP}$ and $\gamma_\mathrm{CP}$.
In this parameterization of the power spectrum the characteristic strain spectrum for the GWB is
\begin{equation}
    h_c(f) = A_\mathrm{CP} \left( \frac{f}{f_\mathrm{yr}} \right)^{(3-\gamma_\mathrm{CP})/2} \,.
    \label{eqn:strain}
\end{equation}
In some cases we fixed $\gamma_\mathrm{CP}=13/3$, equivalent to $\alpha=(3-\gamma_\mathrm{CP})/2 = -2/3$, the expected spectrum for a GWB composed of circular supermassive binary black holes \citep{phi2001}, and in others we left $\gamma_\mathrm{CP}$ as a free parameter.
To determine the number of Fourier frequencies used in the power law CP model, we fit the power spectrum with a broken power law model.
The broken power law is the sum of the standard, red power law and a white spectrum.  This is implemented as a single spectrum with a fixed spectral index at low frequencies that smoothly transitions into a flat, white noise dominated spectrum at high frequencies:
\begin{equation}
    P(f) = \frac{{A_\mathrm{CP}}^2}{12\pi^2} f_\mathrm{yr}^{-3} \left(\frac{f}{f_\mathrm{yr}}\right)^{-\gamma_\mathrm{CP}} \left[ 1 + \left(\frac{f}{f_\mathrm{bend}}\right)^{1/\kappa}\right]^{\kappa\,\gamma_\mathrm{CP}} \,,
    \label{eqn:broken_pl}
\end{equation}
where $f_\mathrm{bend}$ is the frequency where the spectral index of the power spectrum changes and $\kappa$ controls the smoothness of the transition.
In this model $P(f) \sim f^{-\gamma_\mathrm{CP}}$ for $f \ll f_\mathrm{bend}$, and $P(f)$ constant for $f \gg f_\mathrm{bend}$. 
As a verification of our power law models, we performed a free spectral analysis, where the power at each frequency is fit independently rather than being constrained by a particular spectral shape, $P(f)$.

\subsection{Frequentist analyses}
\label{sec:DAmethods.freq}
As a comparison for our primary Bayesian data analysis pipeline, we performed a frequentist analysis using the noise-marginalized \textit{optimal statistic}.
The optimal statistic is an estimator for the amplitude of the GWB based on the inter-pulsar correlations \citep{abc+2009,dfg+13,ccs+2015}.
Its original derivation assumed the pulsars have no intrinsic red noise.
The noise-marginalized optimal statistic uses posterior samples from the Bayesian data analysis to marginalize over the pulsars' red noise.
It has been shown to more accurately estimate the amplitude of the GWB when the pulsars have intrinsic red noise \citep{vit+2018}, as is the case in IPTA DR2.

\section{Results}
\label{sec:results}

The IPTA DR2 data set with its large number of pulsars (53 for this work), long timespan, and various independent observing systems offers a wealth of different analysis opportunities.
Here we present a selection of analyses to give a complete picture of what IPTA DR2 teaches us and how it compares to other PTA data sets.

\subsection{IPTA DR2 data set}

We first show results from the full combined IPTA DR2 data set using methods which differ in their spectral modeling and choice of ORF. The full standard GWB search uses all the available information from the auto- as well as the cross-terms (which are assumed to follow the HD correlation). As the cross-terms, which come from the inter-pulsar correlations, are the defining feature of the GWB, we insist on their presence in order to confidently claim a detection. However, the auto-correlations are initially the dominant source of information, especially for spectral parameter estimation, and the detection of power in them is considered to be the first hint of a GWB \citep{Romano:2020sxq}.
It is important to emphasize that detecting the auto-correlations alone is insufficient to claim a detection of a GWB.

\subsubsection{Common-spectrum process}

\begin{figure}
\includegraphics[width=\columnwidth]{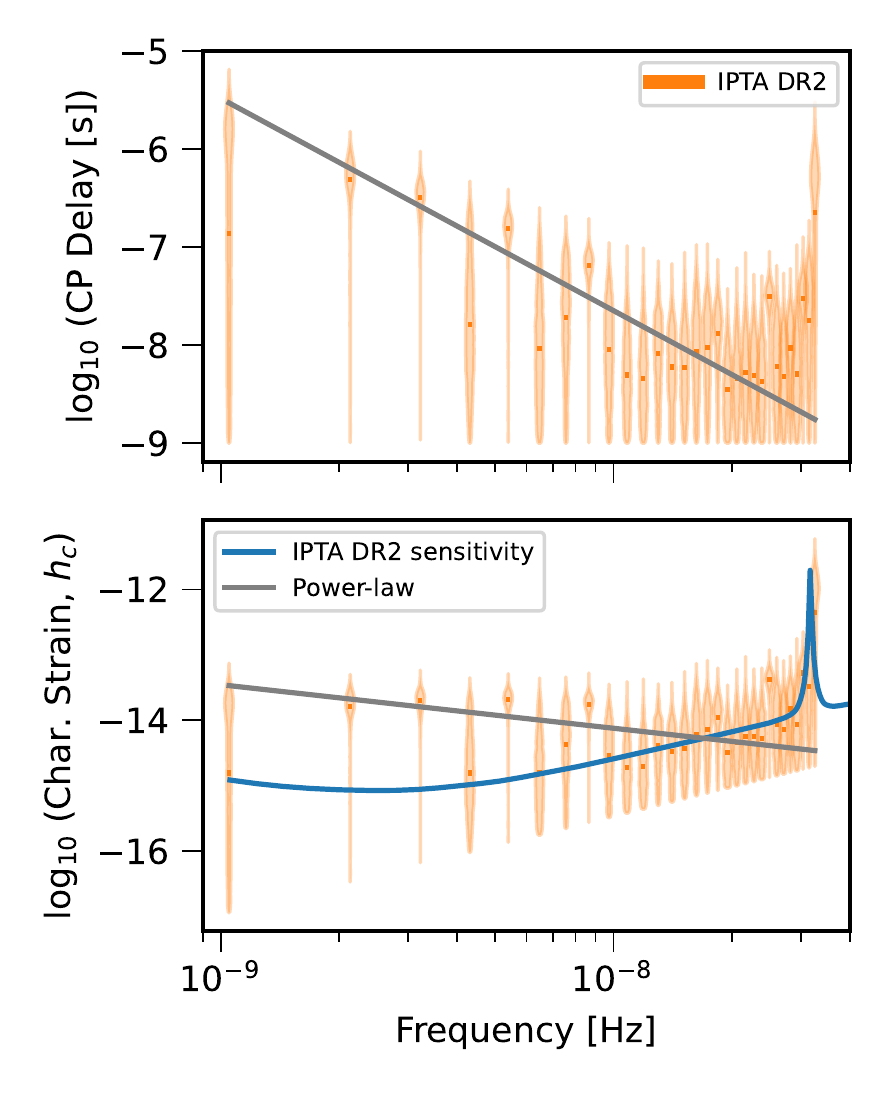}
\caption{IPTA DR2 free spectrum and characteristic strain: The top panel shows the power in terms of time of arrival residuals at  frequencies in the nanohertz band for the full PTA. The maximum likelihood power law is shown overlaid on posteriors for free spectral parameters, a generic model that measures power at various frequencies without imposing any empirical model. The bottom panel shows the power law and free spectral information from above converted into units of characteristic strain, i.e., the noise power measured in the same units as GW amplitude. The additional line shows the characteristic strain for the detector using the noise parameters for the pulsars. The lower limit of all violins is a result of the lower bound of the prior range for each frequency component.}
\label{fig:allFS}
\end{figure}

\begin{figure}
\centering
\includegraphics[width=0.9\columnwidth]{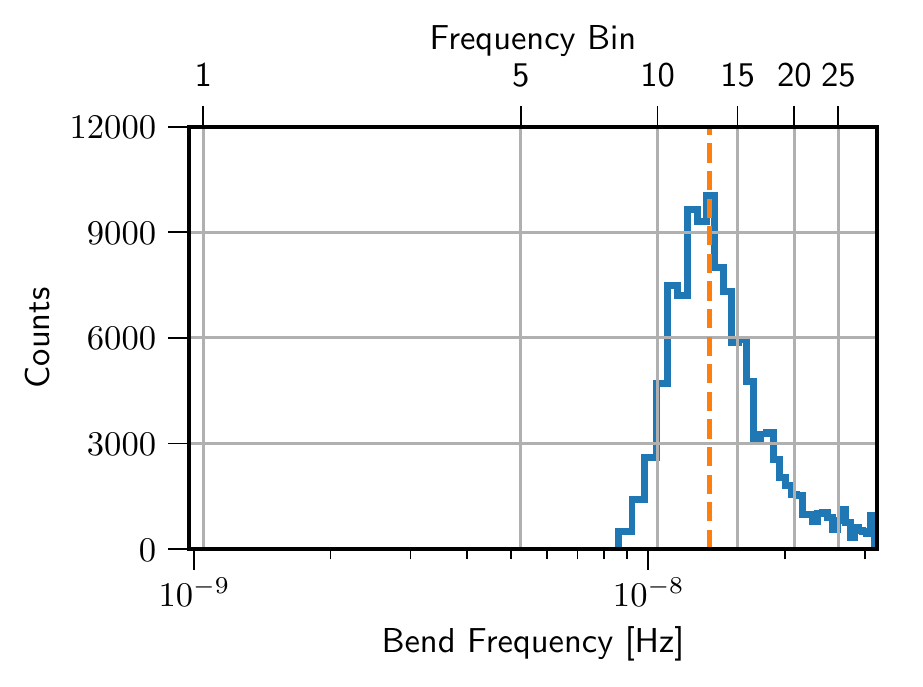}
\caption{Bend Frequency: The posterior for the bend frequency parameter in a broken power law search is shown. The peak of the posterior is at the 13th frequency, $13/T$ for the data set, denoted by the dashed vertical line.}
\label{fig:bend_freq}
\end{figure}

\begin{figure*}
    \centering
   \subfloat[]{\includegraphics[width= 0.5\textwidth]{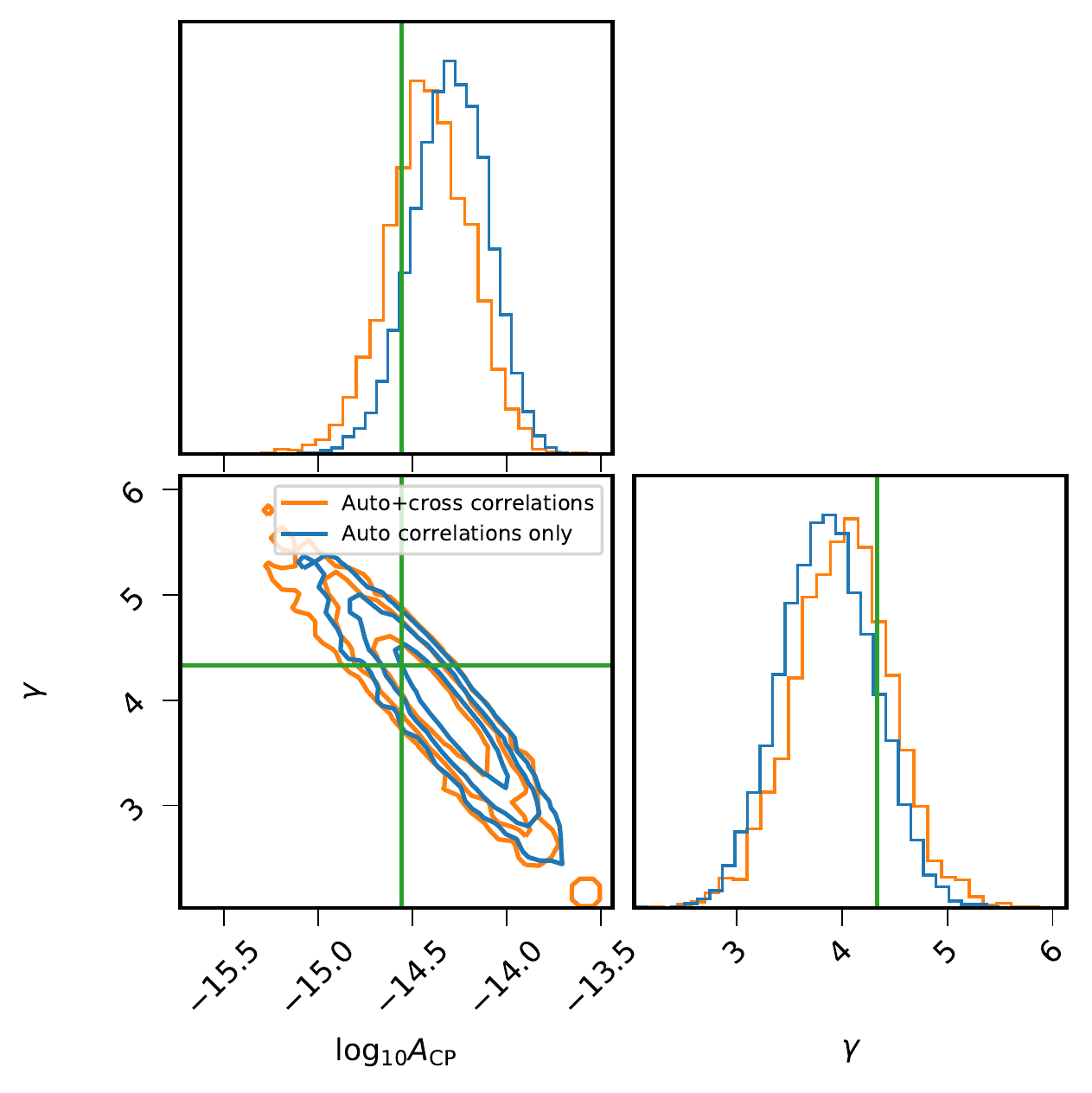}}
    \hfill
    \subfloat[]{\includegraphics[width= \columnwidth]{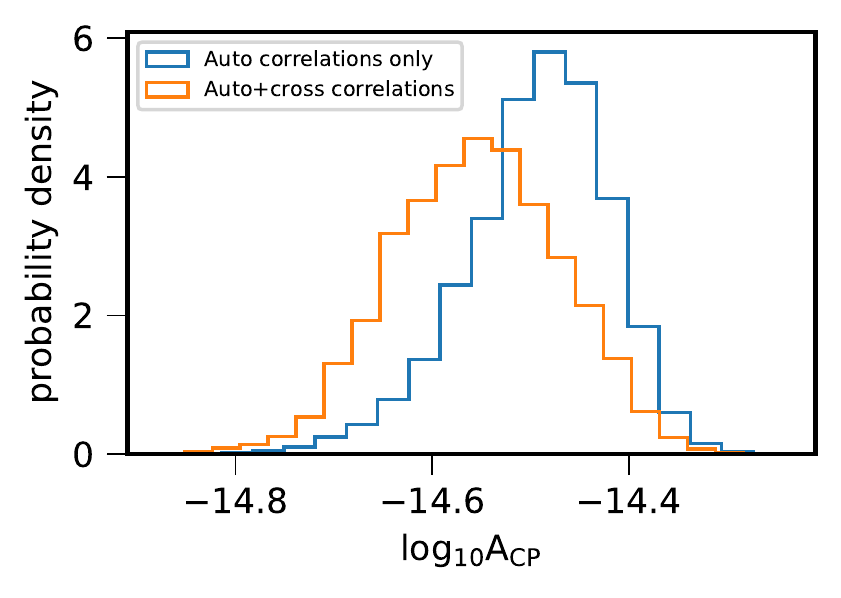}}
    \caption{Comparison of common-spectrum process parameters when using auto-correlations only and the full auto+cross-correlated HD model.
    \textit{left:} 2D posterior for common-spectrum process power law parameters. Green lines mark $\gamma=13/3$ and $A_{\rm CP} = 2.8 \times 10^{-15}$, while the contours represent the 1--, 2-- and 3--$\sigma$ confidence intervals.
    \textit{right:}  1D posterior for common-spectrum process power law amplitude, using fixed spectral index $\gamma=13/3$.}
    \label{fig:varied_gamma}
\end{figure*}

To begin, we apply a free spectral model for the CP, measuring the amount of power at each sampling frequency independently, up to 30 frequency bins, the result of which is shown in the top panel of \autoref{fig:allFS}. The common red noise power can be converted into GW strain using Eqn. \eqref{eqn:strain}. This alternative representation of the IPTA DR2 analysis can be found in the lower panel of \autoref{fig:allFS}. For reference, we include the predicted sensitivity curve, made using \texttt{hasasia}~\citep{hrs2019a,hrs2019b} and the measured white noise parameters of the DR2 data set. Note that the noise power spectral density used in this curve only contains TOA errors, EFAC, EQUAD and ECORR, and does not contain any estimates for the red noise, as for many pulsars it is difficult, and in fact the point of this analysis, to disentangle intrinsic red noise from a GWB. Hence the low-frequency end of the sensitivity  represents a ``best case'' scenario for comparison.
The lowest frequency bin corresponds to the longest timespan, and only two pulsars, J1939+2134 (29 yr) and J1857+0943 (28 yr), have observation baselines sufficient to probe this frequency. However, both have significant RN with spectral indices $\sim3.3$ \citep{pdd+19}.
Therefore, it is not surprising that we do not confidently detect any power there, evidenced by the wide tail extending to low power and median of $\sim 10^{-7}$~s. However, the second, third, fifth, and eighth frequency bins show power well above the expected sensitivity of IPTA DR2. This could either be the emergence of a GWB or some other unmodeled noise process.

The CP power spectrum can be modeled with a simple power law using Eqn. \eqref{eqn:psd}. \cite{abb+2020} have shown that the choice of the number of modeled frequencies can affect the constraints on the power law amplitude and spectral index. Thus, we apply the broken power law model from Eqn. \eqref{eqn:broken_pl} to find the optimal number of frequency bins for the analysis. \autoref{fig:bend_freq} shows the marginalized posterior on the bend frequency. We can identify a clear peak at the 13th frequency in $N/T$, corresponding to a frequency of $1.4 \times 10^{-8}$ Hz, indicated by the orange dashed line. For the remainder of this work, we will limit the search to use only the lowest 13 frequencies with the simple power law model. This produces constraints equivalent to an analysis using the broken power law, but in a simpler and computationally efficient way.

We have also verified that the addition of the \texttt{BAYESEPHEM} SSE model \citep{vts+2020} to the analysis does not change our results significantly from an analysis that fixed the SSE to DE438. The nearly 30 years of timespan allows for the separation of the SSE effects from other correlated signals \citep{vts+2020}. For simplicity, we will only show results with DE438, unless stated otherwise.

\begin{table}
\caption{Bayes factors model comparison: The table shows the logarithmic Bayes factors for a number of model comparisons from the hypermodel and factorized likelihood (marked with an asterisk$^*$) analyses. The preferred model is on the left side of the two models. Brackets indicate the uncertainty in the last digit of the Bayes factors.}
\begin{center}
\begin{tabular}{|c|c|}
\hline
Model comparison & $\log_{10}\rm{BF}$\\
\hline
HD vs CP & 0.3111(6)\\
CP vs Pulsar Noise & 8.2$^*$\\
CP vs Monopole & 4.67(2)\\
CP vs Dipole & 2.28(3)\\
\hline
\end{tabular}
\end{center}
\label{tab:bf}
\end{table}

\autoref{fig:varied_gamma} compares the results when using two different ORFs.
The model that uses only the auto-correlation terms, which we denote CP in \autoref{tab:bf}, is very strongly favored over a model with only intrinsic pulsar noise and no common-spectrum process with $\log_{10}$ Bayes factor of $8.2$.
Despite the large Bayes factor in favor of the CP, this does not suffice to claim a GWB detection, as we have only used the auto-correlations. This strong evidence only indicates that a number of pulsars have red noise with similar spectral characteristics.
We must turn to the cross-correlations to determine if this CP is HD correlated as a GWB should be.
Using the full HD ORF containing both auto- and cross-correlations, we find only middling evidence in favor of the auto+cross HD model.
The $\log_{10}$ Bayes factor for the full HD model compared to the auto-correlated only CP is $0.3$, as shown in \autoref{tab:bf}.

\autoref{fig:varied_gamma}a shows the 2D posterior contours of these two models are in relatively good agreement.
A small shift towards lower amplitudes and higher spectral index can be seen when using the full HD ORF with both auto- and cross-correlations.
Using the auto-correlation terms only, we find $A_{\rm CP} = 5.1^{+6.7}_{-3.1} \times 10^{-15}$ and $\gamma_{\rm CP} = 3.9 \pm 0.9$, where the errors represent 95\% credible regions.
Using the full HD ORF we can constrain the CP power law to $A_{\rm CP} = 3.8^{+6.3}_{-2.5} \times 10^{-15}$ and $\gamma_{\rm CP} = 4.0 \pm 0.9$.

When we fix the power spectrum index to $\gamma = 13/3$ as shown in \autoref{fig:varied_gamma}b, it is clear that full HD model finds a systematically lower amplitude.
In this case we find an amplitude of $A_{\rm CP} = 3.2 \pm 1.0 \times 10^{-15}$ for the auto-correlation only analysis and an amplitude of $A_{\rm CP} = 2.8^{+1.2}_{-0.9} \times 10^{-15}$ using the full HD ORF, where the uncertainties represent the 95\% credible regions.
These results are in broad agreement with published constraints on the CP.
A more detailed comparison can be found in Section \ref{sec:comp_new}.

\begin{figure}
    \centering
    \includegraphics[width=\columnwidth]{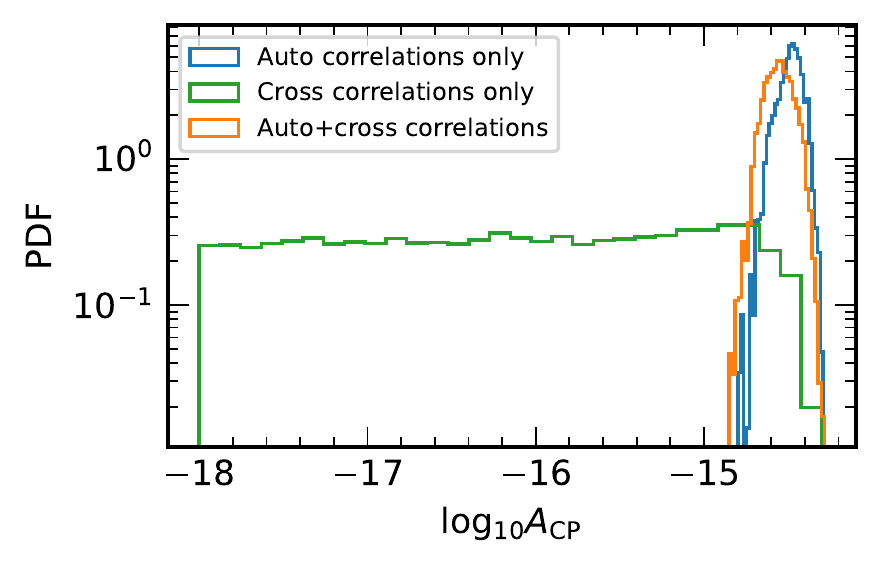}
    \caption{The constraints from the split ORF analysis on the common process amplitude with spectral index fixed to $\gamma = 13/3$. The posterior from the auto-correlations is well constrained, while the posterior from the cross-correlations is unconstrained, but prefers amplitudes slightly smaller than the auto-only analysis. The effect of this is seen in the amplitude posterior from the full auto+cross-correlation model, where the posterior peaks at lower amplitude than the auto-only analysis.}
    \label{fig:split_orf}
\end{figure}

\subsubsection{Split ORF analysis}
Similar to how we may consider the auto-correlation parts of the ORF alone, the full ORF can be split into two independent processes. In this case the auto-correlation and the cross-correlation parts each have their own independent amplitude, as was done in \citet{abb+2020}.
In the HD ORF the auto-correlation part is $\Gamma_{aa}=1$ and the cross-correlation parts are suppressed by at least a factor of 2, $\Gamma_{ab}<0.5$.
This makes the cross-correlations harder to constrain.
\autoref{fig:split_orf} shows the posteriors for the two amplitudes of a split ORF analysis for fixed $\gamma=13/3$, compared to the full auto+cross-correlation model.
The cross-correlations do not have sufficient precision to place constraints on the amplitude of the GWB. However, they do place a 95\% upper limit of $3.6\times 10^{-15}$ on the GWB when the prior choice is taken into account. This is consistent with the amplitude derived using the full auto- and cross-correlation model in Fig.~\ref{fig:split_orf}.
The auto-correlation terms are much more informative.
Combining the information from both shifts the amplitude towards lower values. This shows that the cross-terms can contribute to the full GWB search, even if they provide less information. The auto-correlations are more likely to be affected by intrinsic pulsar noise. Using a more sophisticated noise model for each pulsar can help produce a more robust estimate on the amplitude of any CP.

\begin{figure}
\includegraphics[width=\linewidth]{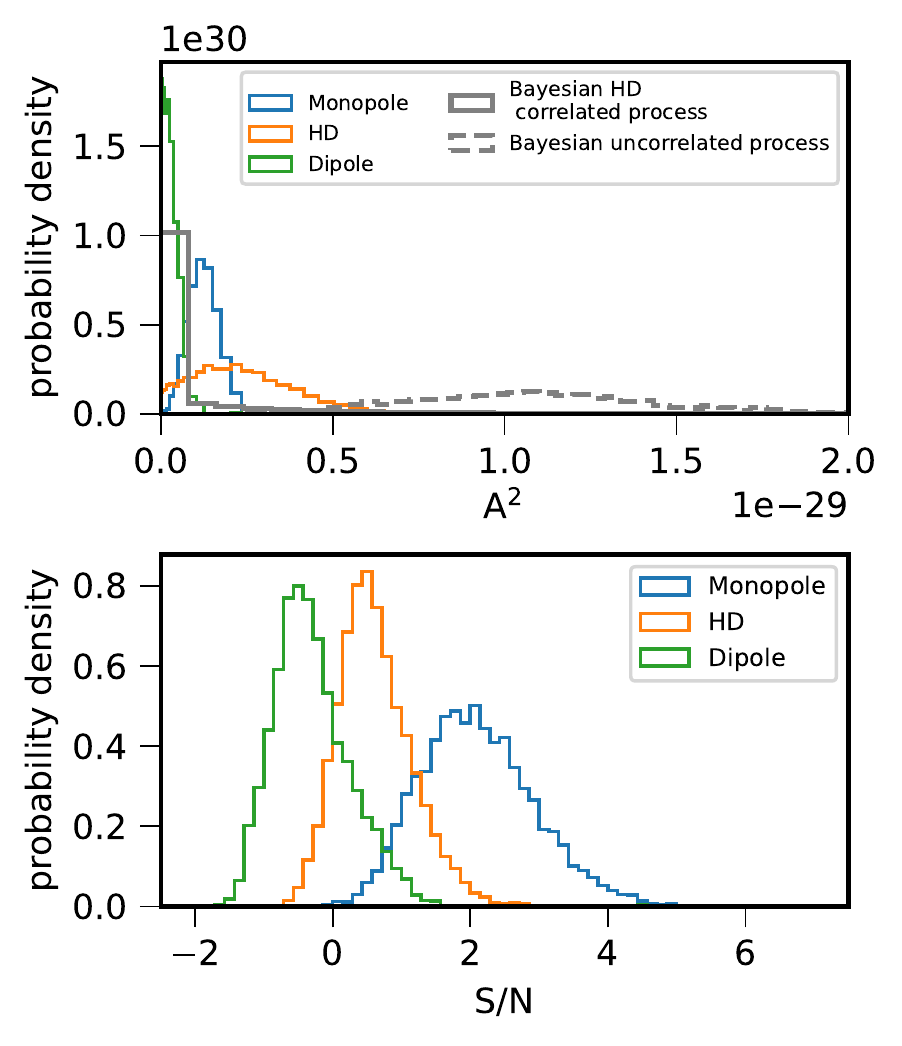}
\caption{Results from the noise marginalized optimal statistic. \textit{Top:} Optimal statistic, $A^2$, distribution for monopole, dipole and HD ORFs. The relevant posteriors from the Bayesian split ORF analysis are also shown for comparison. \textit{Bottom:} The signal-to-noise (S/N) distribution for monopole, dipole and HD ORFs.}
\label{fig:OS}
\end{figure}

\begin{table}
	\caption{The $p$-values calculated from various false alarm analyses of the data set. The measured values of the S/N are compared to the distribution of $10$k analyses where the correlations are broken with phase shifts and sky scrambles. Since a monopolar spatial correlation is uniform across the sky, sky scrambles are unable to break the correlations. Hence only the phase shift $p$-value is quoted.}
	\centering
	\begin{tabular}{|c|c|c|}
	    \hline
	    & $p$, phase shift & $p$, sky scramble \\ 
	    \hline
    	    HD ORF  & 0.25 & 0.24 \\ 
	    \hline
	    Monopole  & 0.03 & $-$ \\ 
	    \hline 
	\end{tabular}
	\label{tab:falsealarm}
\end{table}

\subsubsection{Optimal statistic} \label{sec:os}

\autoref{fig:OS} shows the amplitudes and signal-to-noise ratio (S/N) that are recovered by the pulsar noise marginalized optimal statistic (OS) method, which uses cross-correlations only.
We find no evidence for a dipolar correlated process, as the amplitude and S/N for this model are centered on 0. SSE systematics are expected to manifest at specific frequencies related to the celestial bodies. The IPTA DR2 data set is long enough to probe lower frequencies that should be less affected by SSE errors \citep[][]{vts+2020}. The S/N $= 0.6^{+1.2}_{-0.8}$ for the Hellings-Downs correlation is insufficient to claim a detection. This is consistent with the Bayesian model selection.
The HD amplitude from the OS seems to be in tension with the Bayesian results for the auto-correlated CP, but consistent with the Bayesian results for the full HD model. This strengthens the case that the cross-terms have a significant role to play in parameter estimation as well as detection confidence. Finally, the OS has the largest S/N $= 2.0^{+1.8}_{-1.4}$ for a monopole with a small amplitude. This can be due to the complexity of IPTA DR2 and some amount of unmodeled noise. 

\begin{figure}
\includegraphics[width=\linewidth]{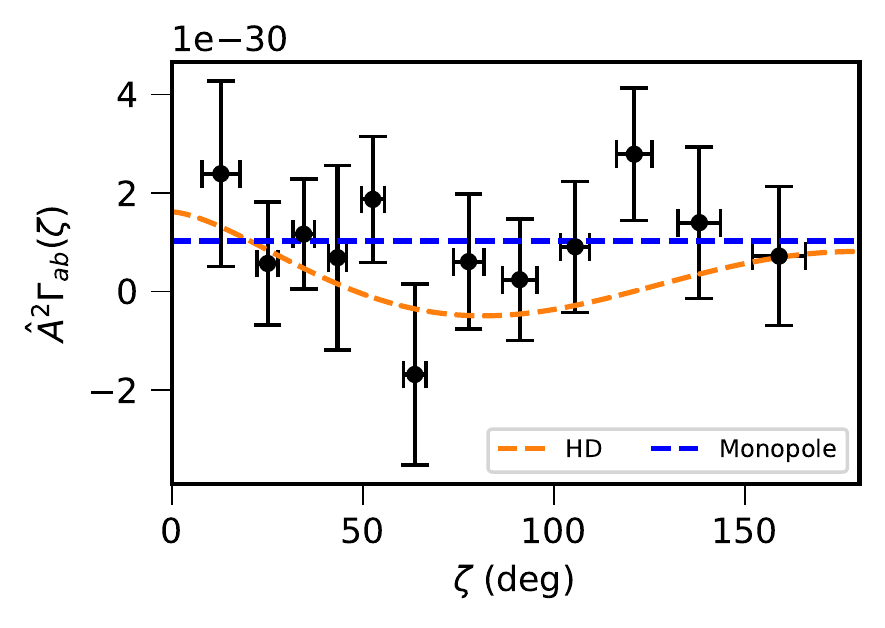}
\caption{Cross-correlation ORF curve from the optimal statistic. The black points indicate the amount of cross-correlation for a given angular separation. Due to the large number of pulsar pairs, we have binned multiple pairs with similar angular separation. The blue and orange dashed lines show the best-fit values for the HD and Monopole correlations.}
\label{fig:correlation}
\end{figure}

As the spatial correlations are not well constrained, see \autoref{fig:correlation}, both the HD and monopolar correlation can fit the data. We have binned pulsar pairs according to their angular separation. Increasing the number of pulsars in the array as well as better timing of pulsars can help to tighten the constraints.

We test the significance of the OS S/N by performing two analyses which estimate the false alarm rate for a given S/N.  So called phase-shifts and sky scrambles \citep{cs2016,taylor+2017} break the correlations between the pulsars leaving the red noise power in the pulsars, but removing evidence for spatial correlations. By analyzing the phase-shifted and sky scrambled data we can determine the rate of observing a particular S/N for a type of correlations in data that has none. These two false alarm studies result in $p$-values, \autoref{tab:falsealarm}, too high to conclude that there is evidence for any correlations. It is possible that the measured HD S/N can therefore arise by chance from a common process with no spatial correlation.

\begin{figure*}
\includegraphics[width=\textwidth]{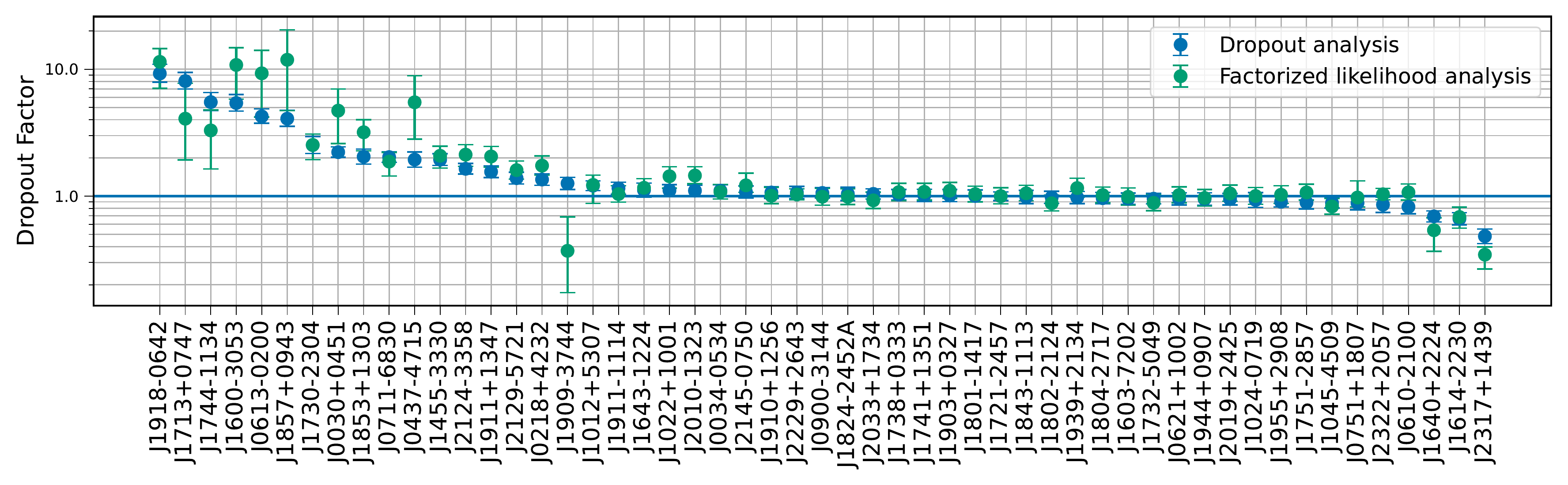}
\caption{Individual pulsar consistency with common-spectrum process, error bars represent 95\% credible intervals. Pulsars with dropout factors $>1$ contribute to the detection of the CP. Dropout factors of $\sim1$ correspond to no evidence for or against the CP, usually due to higher white noise levels and/or shorter observation timespans. Pulsars with dropout factors $<1$ are in tension with the CP.}
\label{fig:dropout}
\end{figure*}

\subsection{IPTA DR2 data subsets}

With the large volume of the IPTA DR2 data set we can look at different subsets to investigate hints of the origin and evolution of the CP signal across different slicings of the full data set.

\subsubsection{Pulsar-based selection}

Since a PTA is made from a number of single pulsars, we can look at how each pulsar contributes to the CP by itself. The dropout factor gives a measure on how consistent a given pulsar's intrinsic red noise is with the CP by comparing a model with and without the CP for the pulsar \citep[see e.g.,][]{abb+2020}. The dropout factors for each pulsar computed using both the traditional hypermodel and factorized likelihood approaches are shown in \autoref{fig:dropout}. About 20 pulsars have factors $> 1$, while only three slightly disfavor the CP, with the remaining pulsars displaying indifference. 

Monte Carlo sampling uncertainties on the dropout factors (computed either way) can be estimated through statistical bootstrapping \citep{efron1994introduction}. In the hypermodel dropout analysis the MCMC chain is re-sampled with replacement, generating a new statistical realization of the sampled chain that is exceptionally unlikely to be identical to the original chain. This process was repeated $10^3$ times, generating many realizations of the MCMC chain from which the dropout factors were computed. Hence a distribution of dropout factors over bootstrap realizations was generated for each pulsar, allowing us to compute median values and 95\% confidence intervals.
A similar procedure was performed for the factorized likelihood approach. For a given bootstrap realization, each individual pulsar's MCMC chain was re-sampled with replacement. With the re-sampled CP posteriors for each pulsar, the factorized-likelihood approach pieces together the dropout factor by iteratively removing pulsars from the array, and all by using bootstrapped pulsar CP posterior chains.
This process was repeated for $10^3$ bootstrap realizations across  $25$ different combinations of meta-parameters used in the factorized-likelihood dropout factor calculation.
The end result is that the median dropout factor and $95\%$ confidence intervals were computed from a total of $2.5\times 10^4$ factorized-likelihood dropout values for each pulsar.
As is seen in \autoref{fig:dropout}, all dropout factors are consistent between both techniques. The vast majority of pulsars have dropout factors with overlapping error bars from both methods, and those that don't are within a few sigma of each other. Those pulsars that show the largest disparity are ones for which there were MCMC sampling inefficiencies that manifested in different stages of the dropout factor calculation, e.g., in the Savage-Dickey density ratio, or in the integral of the $(N-1)$ array's CP likelihood over the posterior of a given pulsar (Taylor et al., in prep.).

The modularity and speed allowed by the factorized likelihood method can be used to approximate different combinations of pulsars within the array. These sub-arrays are a useful way to verify and understand the results we see in the full array. We created four sub-arrays, consisting of pulsars with the highest/lowest dropout factors, longest/shortest timespans. These pulsars were selected by sorting all pulsars in the array by their dropout and time-span characteristics, and then taking the top half of 27 pulsars and the bottom half of 26 pulsars. The Savage-Dickey density ratio was calculated for these sub-arrays to compare them to that of the full array. The sub-array made up of the top half of pulsars according to dropout factor had a Savage-Dickey density ratio of $5.6 \times 10^{9}$, an order of magnitude larger than that of the unaltered array, $1.6 \times 10^{8}$. The corresponding sub-array of the bottom half of pulsars based on dropout had a density ratio of $1.8$. When comparing the pulsars with the longest and shortest timespans, the sub-array with the longer timespans had a Savage-Dickey density ratio of $1.8 \times 10^{7}$ and the shorter timespan sub-array had a density ratio of $3.6$. These results were not surprising, as the dropout factor is a method of measuring the evidence for the array's common process in a particular pulsar, so by removing those that have low (high) dropout factors, the evidence for the CP will increase (decrease). Similarly, pulsars with short timespans are not sensitive to the lowest frequencies explored by the array when in combination with longer timespan pulsars.

\begin{figure*}
\includegraphics[width=\textwidth]{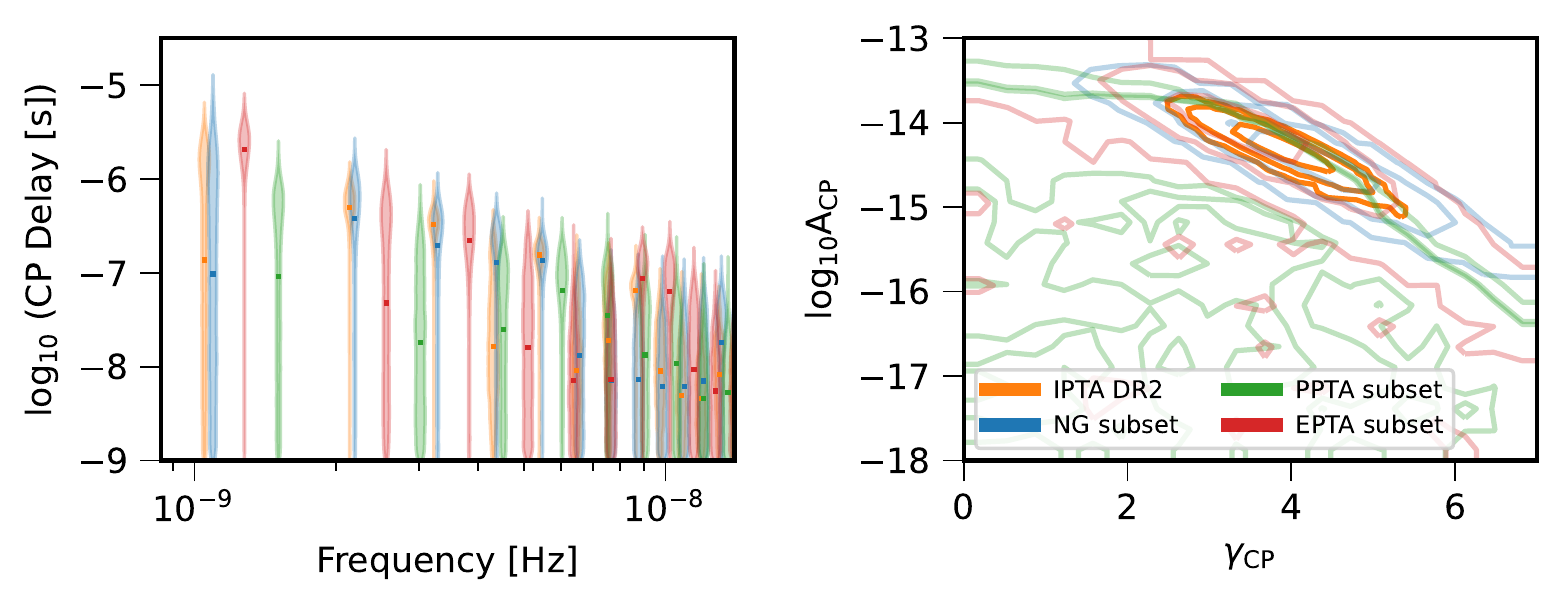}
\caption{Comparison of IPTA DR2 to constituent data sets. \textit{left:} Free spectral common-spectrum process model.  The different sampling frequencies are a result of the constituent subsets covering different timespans.  In all cases the lowest frequency is the inverse observation time. \textit{right:} 2D posterior for CP parameters log-amplitude and spectral index, where the contours represent 1--, 2--, and 3--$\sigma$ confidence intervals.  The combined IPTA DR2 data set is more constraining that its parts.}
\label{fig:constituent}
\end{figure*}

\subsubsection{Splitting IPTA DR2 by time}

To test the evolution of the common-spectrum process, the DR2 is split into two data sets that have equally long timespans, i.e., cutting the DR2 in two time slices \citep[in a similar manner as][]{hst+2020}. The two data sets are not fully equivalent though: the early part contains only 19 pulsars and is mostly dominated by single-radio frequency observations, while the second part has data from all 53 pulsars as well as multi-radio frequency coverage and  higher quality timing measurements. Each data set is then analysed separately. We find that the first half gives little information to the CP, with a broad power law 2D posterior contour that still encompasses the contour from the full data set. The second half contains the majority of information and produces almost identical constraints as the full data set. This is the expected evolution as the quality of the data set gradually improves over time \citep{hst+2020}.

\begin{figure*}
\includegraphics[width=\textwidth]{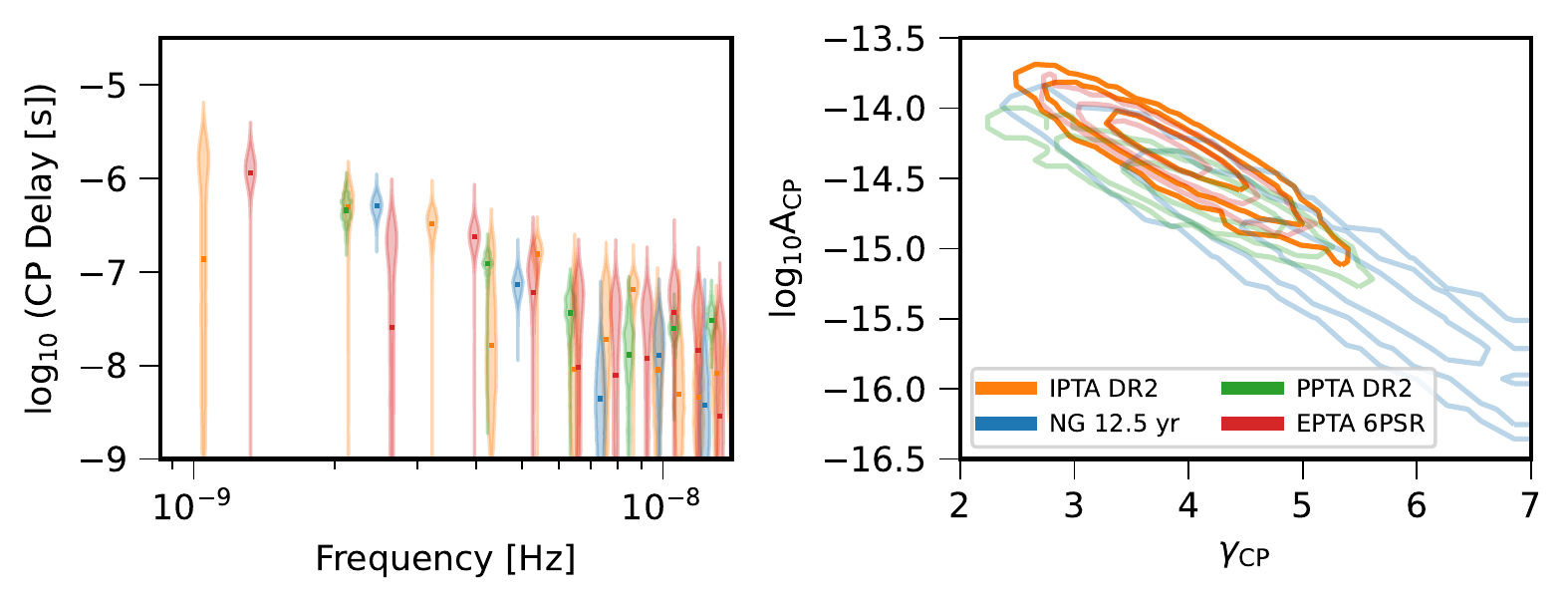}
\caption{Comparison of IPTA DR2 to other recent data sets. \textit{left:} Free spectral common-spectrum process model. The inclusion of legacy data not used in recent PTA analyses allows IPTA DR2 to reach lower frequencies despite missing the most recently collected data. \textit{right:} 2D posterior for CP parameters log-amplitude and spectral index, where the contours represent the 1--, 2--, and 3--$\sigma$ confidence intervals.  All recent data sets are in broad agreement on the characteristics of a common-spectrum process.}
\label{fig:prospective}
\end{figure*}

\begin{figure}
\includegraphics[width=\columnwidth]{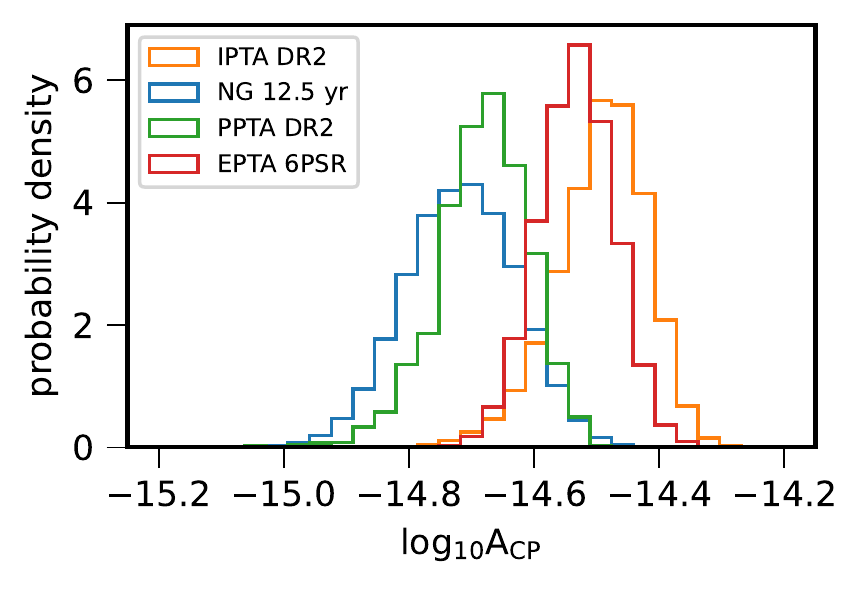}
\caption{CP amplitude posteriors for fixed spectral index, $\gamma=13/3$. IPTA DR2 and EPTA DR2 find a systematically higher amplitude for the common-spectrum process than NANOGrav 12.5 yr and PPTA DR2, although the disagreement is not substantial.}
\label{fig:1Dcompare}
\end{figure}

\subsubsection{Constituent data sets}

We can also select the data that were provided by the constituent PTA collaborations to get three data subsets: EPTA, NANOGrav and PPTA. As each data subset has a different timespan, we set a frequency cutoff at $1.4 \times 10^{-8} {\rm Hz}$ to limit the number of frequencies for the analyses. \autoref{fig:constituent} shows that IPTA DR2 produces the tightest constraints on the CP power law compared to the constituent data sets. While the PPTA data is still consistent with a upper limit, some support for a common red noise can be found with the EPTA and NANOGrav data. The free spectra also show consistency with a power law model that spans across all three constituent data sets and IPTA DR2.

\subsection{Comparison with other recent data sets}
\label{sec:comp_new}

Since the data from the regional PTAs were combined to form the IPTA DR2, the regional PTAs have continued to collect data and improve their data analysis methodology.
We can compare the results using the older IPTA DR2 data set and the most recent data sets from NANOGrav, PPTA and EPTA. Compared to the constituent PTA data sets, the recent NANOGrav data set includes $\sim$ 4 more years and 10 new pulsars, the PPTA expands by $\sim$ 3 years and 7 pulsars, the EPTA DR2 adds $\sim$ 7 years for 6 pulsars \citep{arz+2021,krh+2020,epta_dr2_gwb}.

\begin{table}
	\caption{Mahalanobis distance between CP parameters (log-amplitude and spectral index) for each pair of PTAs.
	For all cases, there is less than 3-sigma separation.}
	\centering
	\begin{tabular}{|c|c|c|c|}
	    \hline
		& \,EPTA\, & \,PPTA\, & NANOGrav \\ \hline
    	IPTA  & 0.6 & 2.6 & 2.6 \\ \hline
	    EPTA  & -- & 2.3 & 2.4 \\ \hline
	    PPTA  & -- & -- & 1.4 \\ \hline
	\end{tabular}
	\label{tab:mahdist}
\end{table}

The published free spectral and power law model recoveries can be found in \autoref{fig:prospective}. For simplicity, we also show the recovered amplitudes at the reference frequency of 1/(1yr) and fixed $\gamma_{\rm CP}=13/3$ in \autoref{fig:1Dcompare}. The Mahalanobis distance $D_M$ acts as a generalization to compute the n-dimensional sigma deviation between two distributions \citep{mahalanobis1936},
\begin{equation}
    D_M = \sqrt{(\vec{\mu_1} - \vec{\mu_2}) \Sigma^{-1} (\vec{\mu_1} - \vec{\mu_2})} \,,
\end{equation}
where $\vec{\mu_{1}}$ and $\vec{\mu_{2}}$ are the mean vectors of the multivariate distributions to be compared and $\Sigma = \Sigma_1 + \Sigma_2$ is the joint covariance.
To quantify the overlap and consistency of the power law parameters as determined using each dataset, the Mahalanobis distance between the 2D posterior distributions are computed in \autoref{tab:mahdist}.
Despite some differences the posteriors overlap better than 3-sigma for all pairs of distributions.

IPTA DR2, using older observations, still shows similar features as the NANOGrav 12.5, 6-pulsar EPTA DR2 and PPTA DR2 analyses, which have added a significant amount of new data to the regional PTA data sets. A future combination of these data sets will boost the total PTA sensitivity in the same way IPTA DR2 is more sensitive than its constituent data sets.
Future combined IPTA data sets will be important for investigating the origin of this common-spectrum process.

\section{Discussion \& outlook}
\label{sec:discuss}

\subsection{Source of the common-spectrum process}

The first IPTA data release did not show signs of a common-spectrum temporally-correlated process, but set an upper limit of $1.7 \times 10^{-15}$ instead. This appears to be in tension with our results from analysis of the second data release with a CP amplitude of $2.8 \times 10^{-15}$. However, there are two major differences to point out: 1) the different choice of priors for the pulsar red, DM and common noise \citep{Hazboun:2020b} and 2) the DR1 upper limit was computed without the use of a SSE uncertainty model \citep{vts+2020}. Both of which have been shown to lead to an increase in the upper limit, alleviating tensions between the DR1 and DR2 CP amplitudes.

As in other recent PTA analyses, we find strong evidence in favor of the CP over the noise only hypothesis. It is important to note that 1) the lack of support for GW-like spatial correlations prohibits any claims of GW detection, however 2) this type of evidence for a similar red noise is expected to precede a detection of spatial correlations \citep{sej+13,Pol:2020igl,Romano:2020sxq}. 

\citet{ppta_dr2_gwb} recently demonstrated that the common-spectrum process model is favored over the noise-only hypothesis when the noise spectra cluster in a similar range, and it is not favored anymore when the noise spectra are drawn from the prior distribution. Because we know that the employed prior distribution for red noise parameters is not representative, it is possible that the evidence we find for a common-spectrum process is caused by a rejection of a null hypothesis rather than by all pulsars exhibiting the spatially-uncorrelated component of a GWB.

Thus, it is important to examine the single pulsar red noise in detail. We have looked at constraints on the simple power law models for the pulsars used in the CP search. In general, pulsars with detectable intrinsic noise have comparable or larger noise than the CP; pulsars without red noise typically have large amount of white noise, such that the CP is `hidden`. One noticeable exception is PSR~J2317+1439, whose noise spectrum falls clearly below the CP, see also its low dropout factor in \autoref{fig:dropout}.

As the search for the common spectrum can be influenced by pulsar intrinsic noise, especially in an inhomogeneous data set, the crucial analysis has to consider information from the cross-correlations. It should be noted that the median amplitudes are slightly different in the analyses with and without spatial correlations, $2.8\times10^{-15}$ vs $3.2\times10^{-15}$. One can also note the stark difference in the posterior for the split ORF analysis, \autoref{fig:split_orf} and the optimal statistic analyses vs the Bayesian uncorrelated analysis, \autoref{fig:OS}. In other analyses \citep{abb+2020,ppta_dr2_gwb,epta_dr2_gwb} the amplitudes between the two analyses are more in line with one another. The difference here could be in part due to the very long baselines of only a handful of pulsars. This legacy data allows only scant opportunity for correlations amongst those few pulsars, while the long baselines allow the detection of auto-correlated power, even in noisier data. Another possibility is that there is unaccounted for noise in individual pulsars that is contaminating the signal. Advanced models to take these pulsar noises into account for the GWB search have been shown to affect the individual pulsar red noise \citep[e.g.,][Chalumeau et al., in prep.]{abb+2020,ppta_dr2_gwb}.

\subsection{Other Correlated Signals}

Spatial correlations in pulsar timing data have been studied in depth in the literature \citep{thk+2018,roebber2019}, and their consideration is an important part of any GW detection procedure. While GWs induce a quadrupole-dominated set of correlations there are other types of spatial correlations between pulsar data sets \citep{rh2017, thk+2018, roebber2019}. Monopolar spatial correlations, i.e., all pulsars seeing the same shifts in residuals irrespective of sky position, can manifest from clock errors, either in the BIPM clock standards, or the various observatory clocks used across the world \citep{hgc+2020}. Dipolar spatial correlations can manifest from the error in measurement of processes  where the motion of the Earth in the solar system is important \citep{cgl+18}. This is most direct in the modeling of the solar system barycenter frame of reference, into which all pulsar TOAs are transformed. Errors in solar wind modeling can also add dipolar correlations \citep{thk+2018}.
While monopole and dipole correlations are theoretically orthogonal to HD correlations in the space of overlap reduction functions, real data with noise can result in some of these modes mixing \citep{roebber2019}.
This mixing could erroneously be detected as a GWB.

The polarization content of the GWB could also deviate from the two tensor transverse (TT) modes predicted by general relativity which lead to the HD spatial correlations \citep{abb+2021b}.
Deviations from general relativity would result in a correlation pattern that differs from HD.
We would like to emphasize that the current data set does not allow us to draw any conclusions on the presence of spatial correlations.
As can be seen in \autoref{fig:correlation} the uncertainties on the spatial correlation coefficients $\Gamma_\text{ab}(\xi)$ determined by the optimal statistic analysis are large. For most angular bins the correlation is indistinguishable from zero, corresponding to the uncorrelated CP.
Close to submission of this paper, we noticed a preprint by \citet{cwh2021}.
They have analyzed the IPTA DR2 searching for alternative GW polarizations and claim evidence for spatial correlations induced by a scalar transverse (ST) polarization mode.
\citet{cwh2021} also report a Bayes factor in favor of HD correlations (TT modes) compared to an uncorrelated CP about six times larger than we find: 12 against our 2.
Even though we have not searched for the ST mode, we would like to highlight that the reported high evidence of spatial correlations in \citet{cwh2021} is contrary to what we conclude using the same data set.
The scalar transverse ORF is positive definite and should be accompanied by positive evidence for a monopolar correlation in analyses using only the cross-terms \citep{abb+2021b}.
This is the case in our optimal statistic analysis where the monopolar correlations have the largest S/N and smallest false alarm $p$-value.
In this sense finding some evidence in favor of ST correlations is not too surprising,
however, we find no conclusive evidence in favor of any correlation pattern. Using more information in our analysis with both auto- and cross-terms disfavors monopolar correlations compared to an uncorrelated CP.
Therefore, the conclusions of \citet{cwh2021} need to be taken with caution.
Finally, \citet{cwh2021} find a Bayes factor in favor of the common process to be several orders of magnitude smaller ($\log_{10}\rm{BF}=4.6$ compared to $8.2$) than we do.
They use a different pulsar noise model, including an additional sinusoidal annual DM variation in all pulsars, which could account for some, but likely not all, of the differences.

\begin{figure*}
\includegraphics[width=\linewidth]{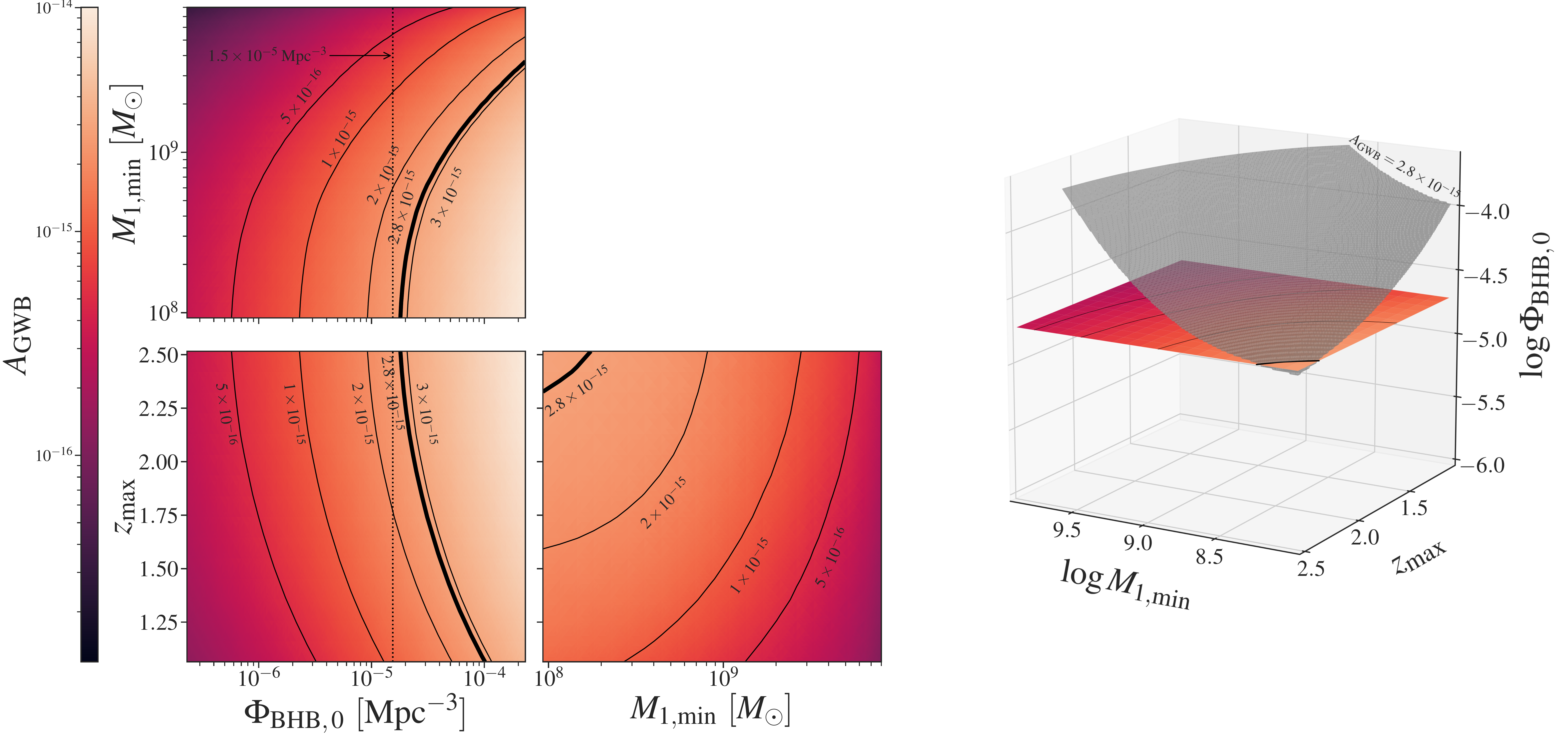}
\caption{The GWB characteristic strain as a function of the local SMBHB number density, $\Phi_{\rm{BHB}, 0}$, and the minimum primary BH mass, $M_{\rm{BH}, 1 \min}$, and maximum redshift, $z_{\rm{max}}$, of the population contributing $\gtrsim 95 \%$ of the GWB signal. {\em Left:} Three representative slices of the strain in this parameter space (one along each axis), with solid contours showing their intersection with isosurfaces of constant strain value ($A_{\rm{CP}}$ shown in bold). {\em Right:} 3D visualization of the $z_{\rm{max}} - M_{\rm{BH}, 1, \min}$ panel from the left and its intersection with an $A_{\rm{GWB}} = 2.8 \times 10^{-15}$ isosurface (gray).}
\label{fig:astro}
\end{figure*}

\begin{figure}
\includegraphics[width=0.45\textwidth]{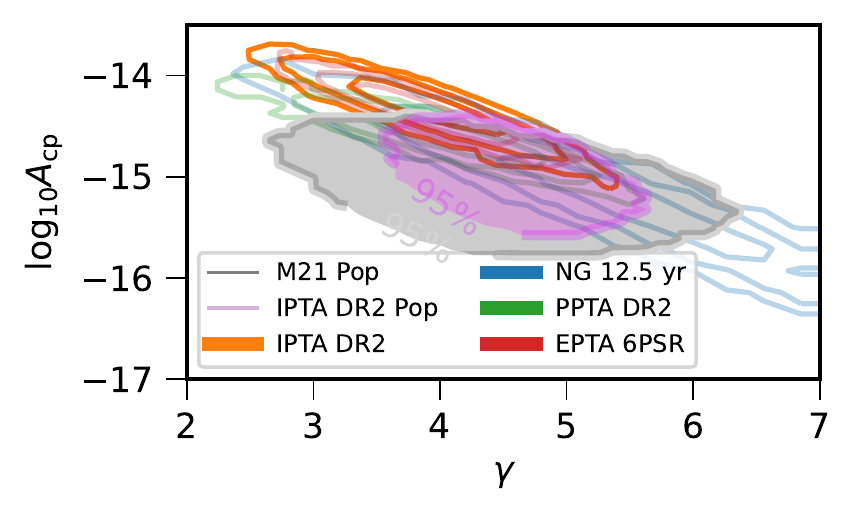}
\caption{Comparison of power law constraints versus theoretical SMBHB populations. The 2D amplitude, spectral index constraints of the CPs from \autoref{fig:prospective} are compared to the region of parameters recovered from a large number of realizations of SMBHB population simulations using astrophysical relations from \citep[M21 Pop,][]{msc+2021} shown in grey contours and this work (IPTA DR2 Pop) shown in purples contours.}
\label{fig:astro_pop}
\end{figure}

\subsection{Astrophysical Interpretation}
\label{sec:astro}

The nHz GWB is generally thought to be dominated by GW emission from SMBHBs \citep{bs+2019}, with the most massive local SMBHBs expected to be individually observable in the next 5-10 years~\citep{mingarelli_local_2017}.
Given that these systems are just a local subset of the cosmological SMBHB population producing the GWB, their local number density, $\Phi_{{\rm BHB}, 0}$, should correlate with the amplitude of the GWB.
The GWB amplitude induced by a cosmological population of circularized SMBHBs can be expressed in geometric units (where $G = c = 1$) as~\citep[e.g.][]{phi2001,svc08,ses13}
\begin{equation}
\label{eq:strain}
     A_{{\rm GWB}}^{2} = \frac{4}{3 \pi^{1 / 3}} \iiint dM_{1} dz dq \frac{\mathcal{M}^{5 / 3}}{\left(1 + z\right)^{1 / 3}} \frac{d^{3}{\Phi_{\rm{BHB}}}}{dM_{1} dz dq},
\end{equation}
where $M_{1}$ is the mass of the primary SMBH, $M_{2}$ the mass of the secondary, $q = M_{2} / M_{1} \leq 1$ is the mass ratio, $\mathcal{M}^{5 / 3} = [q (1 + q)^{-2}] M^{5 / 3}$ is the SMBHB chirp mass with total binary mass $M = M_{1} + M_{2}$, and $d^{3}{\Phi_{\rm{BHB}}} / (dM_{1} dz dq)$ is the differential comoving number density of SMBHBs per unit $M_{1}$, $z$, and $q$.

To determine the local number density of SMBHBs implied by a $A_{\rm{GWB}} \approx 2.8 \times 10^{-15}$ background we use the quasar-based SMBHB model from \citet{casey-clyde_quasarbased_2021}, which assumes proportionality between SMBHB and quasar populations (which may be triggered by galaxy major mergers, \citealt{stemo_catalog_2020}) over mass and redshift. This has the effect of setting $A_{{\rm GWB}} \propto \sqrt{\Phi_{{\rm BHB}, 0}}$, so that $A_{{\rm GWB}}$ directly implies $\Phi_{{\rm BHB}, 0}$.
To check coverage of the entire signal from SMBHBs over mass and redshift, we parameterize $A_{{\rm GWB}} = A_{{\rm GWB}}(\Phi_{\rm{BHB}, 0}, M_{1, \min}, z_{\max})$, where $M_{1, \min}$ and $z_{\max}$ are the minimum primary SMBH mass and maximum redshift, respectively, in \autoref{eq:strain} \citep{casey-clyde_quasarbased_2021}.
We plot this parameterization of the GWB compared to various strain measurements in \autoref{fig:astro}, including an $A_{\rm{GWB}} \approx 2.8 \times 10^{-15}$ signal (bold contour, gray isosurface).
The 2D panels of \autoref{fig:astro} show three representative slices from this parameter space (one along each axis), with contours denoting their intersection with isosurfaces of constant GWB signal amplitude.
The 3D plot shows the bottom right 2D panel in this 3D parameter space, along with its intersection with an $A_{\rm{GWB}} \approx 2.8 \times 10^{-15}$ isosurface.
We find that recovery of a background amplitude like $A_{\rm{CP}}$ requires $\Phi_{\rm{BHB}, 0} \approx 1.5 \times 10^{-5}\;\rm{Mpc}^{-3}$ (corresponding to the bottom right 2D panel in \autoref{fig:astro}), roughly an order of magnitude larger than the $\sim 1.6 \times 10^{-6}\;\rm{Mpc}^{-3}$ number density implied by \citet{mingarelli_local_2017}.

Besides this new quasar-based method the standard approach to determining the local number density $\Phi_{\rm{BHB}, 0}$ is to model $d^{3}{\Phi_{\rm{BHB}}} / (dM_{1} dz dq)$ using major mergers and empirically observed galaxy and black hole relations \citep[e.g.,][]{sbs2016,csc2019,msc+2021}. Following the methods from \cite{msc+2021} we analyze the IPTA DR2 CP amplitude. \autoref{fig:astro_pop} compares the spread of amplitude and spectral index from the IPTA DR2 CP against values recovered from realizations of SMBHB population simulations. The original population constraints using the NANOGrav \citep{abb+2020} frequency bins are shown by the grey shaded area. Repeating the analysis with the frequency coverage of the IPTA DR2 gives the purple shaded contours. As we reach into lower frequencies the simulations become more constrained towards the expected spectral index $\gamma = 13/3$. Limiting the SMBHB chirp mass $\mathcal{M} > 10^{8.5} M_\odot$ in the integral of \autoref{eq:strain} we get $\Phi_{\rm{BHB}, 0} \approx 3.0 \times 10^{-5}\;\rm{Mpc}^{-3}$, which is about a factor of 20 times larger than the number density from \citet{mingarelli_local_2017}.

\subsection{Outlook for other GWBs}
Although we have evidence for a single common process, whose amplitude and spectral index are consistent with predicted values from a population of SMBHBs with a number density $\Phi_0 \approx 10^{-5} {\rm Mpc}^{-3}$, other sources could also be plausible, for example, primordial black holes \citep[e.g.,][]{vv2021,dfr2021,kt2021}, cosmic strings \citep[e.g.,][]{el2021,bbs2021,bow2021} or phase transitions \citep[e.g.,][and references therein]{abb+2021a}. These sources could produce a GWB consistent with the CP contours from PTA data.

\cite{Pol:2020igl} have shown that the initial confident detection of a GWB including HD correlation will place very stringent constraints on the properties of the possible source of GWBs. It is also possible to have several backgrounds affecting the data, splitting the total common power into several components. A detailed study into how well we can separate multi-component GWBs is underway.

\subsection{Modern Noise Mitigation}
The PTA community continues to develop new data analysis strategies towards the detection of gravitational waves in pulsar timing data.
\citet{ng11cw,ng11bwm} showed that unmodeled noise features in a single pulsar could leak into the gravitational wave channel for deterministic continuous GW and GW memory signals respectively.
More recently immense effort was put into the development of individualized noise models for PPTA pulsars, demonstrating that sensitivity can be gained from better modeling \citep{Goncharov:2020krd}.
Similar advanced noise modeling efforts are currently underway in both NANOGrav (Arzoumanian et al., in prep.) and EPTA (Chalumeau et al., in prep.).
More sophisticated noise modeling is important, because many types of noise can add steep spectral index, low frequency power to pulsar data sets, complicating GWB recovery. For example, noise from fluctuations due to the interstellar medium or time-correlated noise from long-term instrumental effects in telescope systems, which could arise from a combination of polarisation miscalibration and secular changes in receiver gain.

\section{Conclusion}
\label{sec:conclusion}

This work shows the immense potential of combining the global efforts of PTA collaborations into one data set. \autoref{fig:allFS} and \autoref{fig:constituent} show that IPTA DR2 is significantly more sensitive than any of the constituent data sets from which it is constructed. While the data in DR2 have now been superseded by more up-to-date efforts, \citep{arz+2021,krh+2020,epta_dr2_gwb}, \autoref{fig:prospective} shows that the sensitivity from combining these older data sets is comparable with these newer single PTA data releases. 

The conclusions of this analysis are broadly similar to the various GWB analyses carried out by the NANOGrav, the PPTA and EPTA \citep{abb+2020,ppta_dr2_gwb,epta_dr2_gwb}. All of these data sets favor the CP model over one with only intrinsic red noise in the individual pulsars. None of these data sets shows clear support for the spatial correlations indicative of a GWB. Therefore a detection of a GWB can not be claimed. The strong detection of red noise that broadly matches the spectral characteristic of a GWB from SMBHBs {\it before} there is support for spatial correlations is expected from our understanding of the change in sensitivity of PTAs \citep{Romano:2020sxq,sej+13}. As was shown in \citep{Pol:2020igl}, {\it if} the power in the auto-correlations of these pulsars is the first sign of the GWB then evidence for the spatial correlations should follow in upcoming data sets. If the next individual PTA data sets show increased support, but are short of detection thresholds, then combining them into an IPTA data set could immediately result in a data set with a significant detection. Such a combination will have the longest timespan, largest number of pulsars and independent observing systems, and will thus enable a robust GWB search.

\section*{Acknowledgements}
\subsection*{Author contributions}
An alphabetical-order author list was used for this paper to recognize the large amount of work and efforts contributed by many people within the IPTA consortium. All authors have contributed to the work via data creation, combination or analysis or contributions to the paper.
The data set was created by a team led by BBPP, MED, PBD, MKe, LL, DJN, SO, SMR and MJK, by combining data sets from the constituent PTAs (EPTA, NANOGrav and PPTA). The data analysis has been the task of the Gravitational Wave Analysis Working Group (WG) of the IPTA. Many analyses have been performed over more than 2 years by PTB, AC, SChe, JAE, JMG, JSH, KI, ADJ, WGL, CMFM, NSP, NKP, DJR, LSc, JS, LSp, SRT and SJV
under the leadership of several WG chairs over this time period: PTB, SChe, JSH, PDL, CMFM, DJR and SRT.

Initial tests, checks of the data quality and exploratory analyses were made by PTB, SChe, JAE, JMG, JSH, KI, CMFM, DJR, JS, SRT and SJV. Although not much of the work is presented in this paper, it helped us tremendously in understanding the subtle details of the data set and built confidence in our analysis. The final results shown come from runs performed by AC, SChe, JSH, ADJ, WGL, NSP, NKP, ASa, LSc, GS, LSp and SRT.
The single pulsar noise analysis was done by SChe. The Bayesian parameter estimation were done by NSP, JSH and WGL.
The optimal statistic analysis was run by JSH and NSP, with help from AC, NKP and LSp for the phase-shifts and sky-scramble null-distributions.
ADJ and JSH have contributed to the Bayes factor computations.
The factorized likelihood method has been applied to the data by LSc and SRT. NSP has computed the dropout factors.
SChe performed the analyses of the constituent data sets, with cross-checks from ASa and GS.
PTB lead the comparison between the IPTA DR2 and recent PTA data sets, which were graciously provided by JS for the NANOGrav 12.5yr, BG for the PPTA DR2 and SChe for the EPTA DR2.
The astrophysical interpretation was lead by JACC and CMFM with contributions from SChe and ASe.

SChe lead the coordination of the paper writing with the help of JSH and PTB.
PTB, JACC, SChe, MED, BG, JSH, CMFM, BBPP, NSP, LSc, RMS, SRT and SJV contributed to the writing of the paper. The figures were created by NSP, JSH, JACC, and SChe.

\subsection*{Acknowledgements}

The International Pulsar Timing Array (IPTA) is a consortium of existing Pulsar Timing Array collaborations, namely, the European Pulsar Timing Array (EPTA), North American Nanohertz Observatory for Gravitational Waves (NANOGrav), Parkes Pulsar Timing Array (PPTA) and the recent addition of the Indian Pulsar Timing Array (InPTA). Observing collaborations from China and South Africa are also part of the IPTA.

The EPTA is a collaboration between European and partner institutes with the aim to provide high precision pulsar timing to work towards the direct detection of low-frequency gravitational waves. An Advanced Grant of the European Research Council to implement the Large European Array for Pulsars (LEAP) also provides funding. Part of this work is based on observations with the 100-m telescope of the Max-Planck-Institut f\"{u}r Radioastronomie (MPIfR) at Effelsberg in Germany. Pulsar research at the Jodrell Bank Centre for Astrophysics and the observations using the Lovell Telescope are supported by a Consolidated Grant (ST/T000414/1) from the UK's Science and Technology Facilities Council. The Nan{\c c}ay radio Observatory is operated by the Paris Observatory, associated to the French Centre National de la Recherche Scientifique (CNRS), and partially supported by the Region Centre in France. We acknowledge financial support from ``Programme National de Cosmologie and Galaxies'' (PNCG), and ``Programme National Hautes Energies'' (PNHE) funded by CNRS/INSU-IN2P3-INP, CEA and CNES, France. We acknowledge financial support from Agence Nationale de la Recherche (ANR-18-CE31-0015), France. The Westerbork Synthesis Radio Telescope is operated by the Netherlands Institute for Radio Astronomy (ASTRON) with support from the Netherlands Foundation for Scientific Research (NWO). The Sardinia Radio Telescope (SRT) is funded by the Department of University and Research (MIUR), the Italian Space Agency (ASI), and the Autonomous Region of Sardinia (RAS) and is operated as National Facility by the National Institute for Astrophysics (INAF).

The NANOGrav Physics Frontiers Center is supported by NSF award number 1430284. The Green Bank Observatory is a facility of the National Science Foundation operated under cooperative agreement by Associated Universities, Inc. The Arecibo Observatory is a facility of the National Science Foundation operated under cooperative agreement by the University of Central Florida in alliance with Yang Enterprises, Inc. and Universidad Metropolitana.

The Parkes radio telescope (Murriyang) is part of the Australia Telescope which is funded by the Commonwealth Government for operation as a National Facility managed by CSIRO.

JA acknowleges support by the Stavros Niarchos Foundation (SNF) and the Hellenic Foundation for Research and Innovation (H.F.R.I.) under the 2nd Call of ``Science and Society'' Action Always strive for excellence -- ``Theodoros Papazoglou'' (Project Number: 01431). SBS acknowledges generous support by the NSF through grant AST-1815664. The work is supported by National SKA program of China 2020SKA0120100, Max-Planck Partner Group, NSFC 11690024, CAS Cultivation Project for FAST Scientific. JACC was supported in part by NASA CT Space Grant PTE Federal Award Number 80NSSC20M0129. CMFM and JACC are also supported by the National Science Foundation's NANOGrav Physics Frontier Center, Award Number 2020265. AC acknowledges support from the Paris \^{I}le-de-France Region. Support for HTC was provided by NASA through the NASA Hubble Fellowship Program grant HST-HF2-51453.001 awarded by the Space Telescope Science Institute, which is operated by the Association of Universities for Research in Astronomy, Inc., for NASA, under contract NAS5-26555. SD is the recipient of an Australian Research Council Discovery Early Career Award (DE210101738) funded by the Australian Government. GD, RK and MKr acknowledge support from European Research Council (ERC) Synergy Grant ``BlackHoleCam'' Grant Agreement Number 610058 and ERC Advanced Grant ``LEAP'' Grant Agreement Number 337062. TD is supported by the NSF AAG award number 2009468. ECF is supported by NASA under award number 80GSFC17M0002.002. BG is supported by the Italian Ministry of Education, University and Research within the PRIN 2017 Research Program Framework, n. 2017SYRTCN. Portions of this work performed at the Naval Research Laboratory is supported by NASA and ONR 6.1 basic research funding. MTL acknowledges support received from NSF AAG award number 200968. Part of this research was carried out at the Jet Propulsion Laboratory, California Institute of Technology, under a contract with the National Aeronautics and Space Administration. JWMK is a CITA Postdoctoral Fellow: This work was supported by the Natural Sciences and Engineering Research Council of Canada (NSERC), (funding reference CITA 490888-16). ASa, ASe and GS acknowledge financial support provided under the European Union's H2020 ERC Consolidator Grant ``Binary Massive Black Hole Astrophysic'' (B Massive, Grant Agreement: 818691). RMS acknowledges support through Australian Research Council Future Fellowship FT190100155. JS acknowledges support from the JPL R\&TD program. This research was funded partially by the Australian Government through the Australian Research Council (ARC), grants CE170100004 (OzGrav) and FL150100148. Pulsar research at UBC is supported by an NSERC Discovery Grant and by the Canadian Institute for Advanced Research. SRT acknowledges support from NSF grants AST-2007993 and PHY-2020265. SRT also acknowledges support from the Vanderbilt University College of Arts \& Science Dean's Faculty Fellowship program. AV acknowledges the support of the Royal Society and Wolfson Foundation. JPWV acknowledges support by the Deutsche Forschungsgemeinschaft (DFG) through the Heisenberg programme (Project No. 433075039).

\section*{Data availability}
The timing data used can be found on \url{https://gitlab.com/IPTA/DR2}. Selected chain files for the analyses presented in this paper can be found on \url{https://ipta4gw.org} and \url{https://doi.org/10.5281/zenodo.5787557}.

\bibliographystyle{mnras}
\bibliography{psrrefs,modrefs,journals}

\ \\
$^{1}$ Institute of Astrophysics, FORTH, N. Plastira 100, 70013, Heraklion, Greece\\
$^{2}$ Max-Planck-Institut f{\"u}r Radioastronomie, Auf dem H{\"u}gel 69, 53121 Bonn, Germany\\
$^{3}$ Argelander Institut f{\"u}r Astronomie, Auf dem H{\"u}gel 71, 53117, Bonn, Germany\\
$^{4}$ X-Ray Astrophysics Laboratory, NASA Goddard Space Flight Center, Code 662, Greenbelt, MD 20771, USA\\
$^{5}$ Universit{\'e} de Paris, CNRS, Astroparticule et Cosmologie, 75013 Paris, France\\
$^{6}$ Moscow Institute of Physics and Technology, Dolgoprudny, Moscow region, Russia\\
$^{7}$ Centre for Astrophysics and Supercomputing, Swinburne University of Technology, Hawthorn, VIC 3122, Australia\\
$^{8}$ Australian Research Council Centre for Excellence for Gravitational Wave Discovery (OzGrav)\\
$^{9}$ Fakult{\"a}t f{\"u}r Physik, Universit{\"a}t Bielefeld, Postfach 100131, 33501 Bielefeld, Germany\\
$^{10}$ Department of Physics and Astronomy, Widener University, One University Place, Chester, PA 19013, USA\\
$^{11}$ ASTRON, Netherlands Institute for Radio Astronomy, Oude Hoogeveensedijk 4, 7991 PD, Dwingeloo, The Netherlands\\
$^{12}$ Department of Physics, Montana State University, Bozeman, MT 59717, USA\\
$^{13}$ Laboratoire de Physique et Chimie de l'Environnement et de l'Espace LPC2E UMR7328, Universit{\'e} d'Orl{\'e}ans, CNRS, 45071 Orl{\'e}ans, France\\
$^{14}$ Station de Radioastronomie de Nan\c{c}ay, Observatoire de Paris, PSL University, CNRS, Universit{\'e} d'Orl{\'e}ans, 18330 Nan\c{c}ay, France\\
$^{15}$ Dipartimento di Fisica ``G. Occhialini", Universit{\`a} degli Studi di Milano-Bicocca, Piazza della Scienza 3, 20126 Milano, Italy\\
$^{16}$ INFN, Sezione di Milano-Bicocca, Piazza della Scienza 3, 20126 Milano, Italy\\
$^{17}$ Cornell Center for Astrophysics and Planetary Science and Department of Astronomy, Cornell University, Ithaca, NY 14853, USA\\
$^{18}$ Department of Physics and Astronomy, West Virginia University, P.O. Box 6315, Morgantown, WV 26506, USA\\
$^{19}$ Center for Gravitational Waves and Cosmology, West Virginia University, Chestnut Ridge Research Building, Morgantown, WV 26505, USA\\
$^{20}$ INAF - Osservatorio Astronomico di Cagliari, via della Scienza 5, 09047 Selargius (CA), Italy\\
$^{21}$ Canadian Institute for Advanced Research, CIFAR Azrieli Global Scholar, MaRS Centre West Tower, 661 University Ave. Suite 505, Toronto ON M5G 1M1, Canada\\
$^{22}$ Kavli Institute for Astronomy and Astrophysics, Peking University, Beijing 100871, P. R. China\\
$^{23}$ Department of Physics, University of Connecticut, 196 Auditorium Road, U-3046, Storrs, CT 06269-3046, USA\\
$^{24}$ Department of Physics and Astronomy, Vanderbilt University, 2301 Vanderbilt Place, Nashville, TN 37235, USA\\
$^{25}$ Department of Physics and Astronomy, Franklin \& Marshall College, P.O. Box 3003, Lancaster, PA 17604, USA\\
$^{26}$ NASA Hubble Fellowship Program Einstein Postdoctoral Fellow\\
$^{27}$ Department of Physics and Astronomy, University of British Columbia, 6224 Agricultural Road, Vancouver, BC V6T 1Z1, Canada\\
$^{28}$ School of Science, Western Sydney University, Locked Bag 1797, Penrith South DC, NSW 2751, Australia\\
$^{29}$ George Mason University, resident at the Naval Research Laboratory, Washington, DC 20375, USA\\
$^{30}$ National Radio Astronomy Observatory, 1003 Lopezville Rd., Socorro, NM 87801, USA\\
$^{31}$ LESIA, Observatoire de Paris, Universit{\'e} PSL, CNRS, Sorbonne Universit{\'e}, Universit{\'e} de Paris, 5 place Jules Janssen, 92195 Meudon, France\\
$^{32}$ Department of Physics, Hillsdale College, 33 E. College Street, Hillsdale, MI 49242, USA\\
$^{33}$ Eureka Scientific, Inc., 2452 Delmer Street, Suite 100, Oakland, CA 94602-3017, USA\\
$^{34}$ School of Physics and Astronomy, Rochester Institute of Technology, Rochester, NY 14623, USA\\
$^{35}$ Department of Astronomy, University of Maryland, College Park, MD 20742, USA\\
$^{36}$ Center for Research and Exploration in Space Science and Technology, NASA/GSFC, Greenbelt, MD 20771, USA\\
$^{37}$ NASA Goddard Space Flight Center, Greenbelt, MD 20771, USA\\
$^{38}$ Department of Physics and Astronomy, West Virginia University, PO Box 6315, Morgantown, WV 26506, USA\\
$^{39}$ Max Planck Institute for Gravitational Physics (Albert Einstein Institute), Am Mu{\"u}hlenberg 1, 14476 Potsdam, Germany\\
$^{40}$ Gran Sasso Science Institute (GSSI), 67100 L'Aquila, Italy\\
$^{41}$ INFN, Laboratori Nazionali del Gran Sasso, 67100 Assergi, Italy\\
$^{42}$ Physical Sciences Division, University of Washington Bothell, 18115 Campus Way NE, Bothell, WA 98011, USA\\
$^{43}$ Australia Telescope National Facility, CSIRO Space and Astronomy, P.O. Box 76, Epping, NSW 1710, Australia\\
$^{44}$ Center for Gravitation, Cosmology and Astrophysics, Department of Physics, University of Wisconsin-Milwaukee, P.O. Box 413, Milwaukee, WI 53201, USA\\
$^{45}$ Department of Astrophysics/IMAPP, Radboud University Nijmegen, P.O. Box 9010, 6500 GL Nijmegen, The Netherlands\\
$^{46}$ Jodrell Bank Centre for Astrophysics, Department of Physics and Astronomy, University of Manchester, Manchester M13 9PL, UK\\
$^{47}$ Center for Interdisciplinary Exploration and Research in Astrophysics (CIERA), Northwestern University, Evanston, IL 60208, USA\\
$^{48}$ Space Science Division, Naval Research Laboratory, Washington, DC 20375, USA\\
$^{49}$ Laboratory for Multiwavelength Astrophysics, Rochester Institute of Technology, Rochester, NY 14623, USA\\
$^{50}$ Jet Propulsion Laboratory, California Institute of Technology, 4800 Oak Grove Drive, Pasadena, CA 91109, USA\\
$^{51}$ National Astronomical Observatories, Chinese Academy of Sciences, Beijing 100101, P. R. China\\
$^{52}$ Astrophysics Group, Cavendish Laboratory, JJ Thomson Avenue, Cambridge, CB3 0HE, UK\\
$^{53}$ Department of Astronomy \& Astrophysics, University of Toronto, 50 Saint George Street, Toronto, ON M5S 3H4, Canada\\
$^{54}$ Green Bank Observatory, P.O. Box 2, Green Bank, WV 24944, USA\\
$^{55}$ Department of Physics, University of the Pacific, 3601 Pacific Ave., Stockton, CA 95211, USA\\
$^{56}$ Canadian Institute for Theoretical Astrophysics, University of Toronto, 60 St. George Street, Toronto, ON M5S 3H8, Canada\\
$^{57}$ Center for Computational Astrophysics, Flatiron Institute, 162 5th Avenue, New York, NY 10010, USA\\
$^{58}$ Dunlap Institute for Astronomy and Astrophysics, University of Toronto, 50 Saint George Street, Toronto, ON M5S 3H4, Canada\\
$^{59}$ Department of Physics, Lafayette College, Easton, PA 18042, USA\\
$^{60}$ Manly Astrophysics, 15/41-42 East Esplanade, Manly, NSW 2095, Australia\\
$^{61}$ Institute of Physics, E{\"o}tv{\"o}s Lor{\'a}nd University, P{\'a}zm{\'a}ny P.s. 1/A, 1117 Budapest, Hungary\\
$^{62}$ Arecibo Observatory, HC3 Box 53995, Arecibo, PR 00612, USA\\
$^{63}$ Universit{\`a} di Cagliari, Dipartimento di Fisica, S.P. Monserrato-Sestu Km 0,700 - 09042 Monserrato (CA), Italy\\
$^{64}$ National Radio Astronomy Observatory, 520 Edgemont Road, Charlottesville, VA 22903, USA\\
$^{65}$ U.S. Naval Research Laboratory, Washington, DC 20375, USA\\
$^{66}$ CSIRO Scientific Computing Services, Australian Technology Park, Locked Bag 9013, Alexandria, NSW 1435, Australia\\
$^{67}$ Australia Telescope National Facility, CSIRO Space and Astronomy, P.O. Box 276, Parkes, NSW 2870, Australia\\
$^{68}$ Theoretical Physics Department, CERN, 1211 Geneva 23, Switzerland\\
$^{69}$ Department of Physics, Oregon State University, Corvallis, OR 97331, USA\\
$^{70}$ Department of Astrophysical and Planetary Sciences, University of Colorado, Boulder, CO 80309, USA\\
$^{71}$ Department of Physics and Astronomy, Swarthmore College, Swarthmore, PA 19081, USA\\
$^{72}$ Department of Physics and Astronomy, Oberlin College, Oberlin, OH 44074, USA\\
$^{73}$ Laboratoire Univers et Th{\'e}ories LUTh, Observatoire de Paris, Universit{\'e} PSL, CNRS, Universit{\'e} de Paris, 92190 Meudon, France\\
$^{74}$ Theoretical AstroPhysics Including Relativity (TAPIR), MC 350-17, California Institute of Technology, Pasadena, CA 91125, USA\\
$^{75}$ Institute for Gravitational Wave Astronomy and School of Physics and Astronomy, University of Birmingham, Edgbaston, Birmingham B15 2TT, UK\\
$^{76}$ Xinjiang Astronomical Observatory, Chinese Academy of Sciences, 150 Science 1-Street, Urumqi, Xinjiang 830011, P. R. China\\
$^{77}$ Purple Mountain Observatory, Chinese Academy of Sciences, Nanjing 210023, P. R. China\\
$^{78}$ Advanced Institute of Natural Sciences, Beijing Normal University at Zhuhai 519087, P. R. China\\

\label{lastpage}

\end{document}